\documentclass[twocolumn, tighten, dvipsnames, floatfix]{aastex63}

\usepackage[]{xcolor}
\usepackage{booktabs}
\usepackage{physics}
\usepackage{longtable}
\usepackage{array}
\usepackage{amsmath}
\usepackage{changepage}
\usepackage{lineno}
\usepackage[flushmargin]{footmisc}
\usepackage{afterpage}
\usepackage{subfigure}

\newcommand{\ret}[1]{\textbf{\textcolor{Bittersweet}{#1}}}
\newcommand{\gui}[1]{\textbf{\textcolor{red}{#1}}}
\newcommand{\azul}[1]{\textcolor{blue}{#1}}
\newcommand{\rafa}[1]{\textbf{\textcolor{magenta}{#1}}}
\newcommand{\helio}[1]{\textcolor{teal}{#1}}
\newcommand{\silvia}[1]{\textcolor{green}{#1}}
\newcommand{\angeles}[1]{\textcolor{purple}{#1}}

\newcommand{\vdag}{(v)^\dagger}
\newcommand\aastex{AAS\TeX}
\newcommand\latex{La\TeX}

\def\arcsec{\hbox{$^{\hbox{\rlap{\hbox{\lower4pt\hbox{$\,\prime\prime$}}
          }}}$} \ }

\def\arcmin{\hbox{$^{\hbox{\rlap{\hbox{\lower4pt\hbox{$\;\prime$}}
          }\hbox{$\frown$}}}$}}

\usepackage{lipsum}
\usepackage{float}

\shorttitle{Targeting Bright Metal-poor Stars in the Milky Way}
\shortauthors{Limberg et al.}

\begin{document}

\title{Targeting Bright Metal-poor Stars in the Disk and Halo Systems of the Galaxy}

\correspondingauthor{Guilherme Limberg}
\email{guilherme.limberg@usp.br}

\author[0000-0002-9269-8287]{Guilherme Limberg}
\affil{Universidade de S\~ao Paulo, Instituto de Astronomia, Geof\'isica e Ci\^encias Atmosf\'ericas, Departamento de Astronomia, \\ SP 05508-090, S\~ao Paulo, Brazil}

\author[0000-0002-7529-1442]{Rafael M. Santucci}
\affiliation{Universidade Federal de Goi\'as, Instituto de Estudos Socioambientais, Planet\'ario, Goi\^ania, GO 74055-140, Brazil}
\affiliation{Universidade Federal de Goi\'as, Campus Samambaia, Instituto de F\'isica, Goi\^ania, GO 74001-970, Brazil}

\author[0000-0001-7479-5756]{Silvia Rossi}
\affil{Universidade de S\~ao Paulo, Instituto de Astronomia, Geof\'isica e Ci\^encias Atmosf\'ericas, Departamento de Astronomia, \\ SP 05508-090, S\~ao Paulo, Brazil}

\author[0000-0001-9723-6121]{Derek Shank}
\affiliation{Department of Physics and JINA Center for the Evolution of the Elements, University of Notre Dame, Notre Dame, IN 46556, USA}

\author[0000-0003-4479-1265]{Vinicius M. Placco}
\affiliation{Community Science and Data Center/NSF’s NOIRLab, 950 N. Cherry Ave., Tucson, AZ 85719, USA}

\author[0000-0003-4573-6233]{Timothy C. Beers}
\affiliation{Department of Physics and JINA Center for the Evolution of the Elements, University of Notre Dame, Notre Dame, IN 46556, USA}

\author[0000-0001-5761-6779]{Kevin C. Schlaufman}
\affiliation{Department of Physics and Astronomy
Johns Hopkins University
3400 North Charles Street
Baltimore, MD 21218, USA}

\author[0000-0003-0174-0564]{Andrew R. Casey}
\affiliation{School of Physics \& Astronomy, Monash University, Wellington Road, Clayton 3800, Victoria, Australia}
\affiliation{ARC Centre of Excellence for All Sky Astrophysics in 3 Dimensions (ASTRO 3D), Canberra, ACT 2611, Australia}

\author[0000-0002-0537-4146]{H\'elio D. Perottoni}
\affil{Universidade de S\~ao Paulo, Instituto de Astronomia, Geof\'isica e Ci\^encias Atmosf\'ericas, Departamento de Astronomia, \\ SP 05508-090, S\~ao Paulo, Brazil}

\author[0000-0003-0852-9606]{Young Sun Lee}
\affiliation{Department of Astronomy and Space Science, Chungnam National University, Daejeon 34134, Republic of Korea}

\defcitealias{placco2019}{P19}
\defcitealias{placco2018}{P18}

\begin{abstract}

We present the results of spectroscopic follow-up for 1897 low-metallicity star candidates, selected from the Best \& Brightest (B\&B) Survey, carried out with the GMOS-N/S (Gemini North/South telescopes) and Goodman (SOAR Telescope) spectrographs. From these low-resolution ($R \sim 2000$) spectra, we estimate stellar atmospheric parameters, as well as carbon and magnesium abundance ratios. We confirm that $56\%$ of our program stars are metal-poor ([Fe/H] $< -1.0$), $30\%$ are very metal-poor (VMP; [Fe/H] $< -2.0$) and $2\%$ are extremely metal-poor (EMP; [Fe/H] $< -3.0$). There are 191 carbon-enhanced metal-poor (CEMP) stars, resulting in CEMP fractions of $19\%$ and $43\%$ for the VMP and EMP regimes, respectively.  A total of 94 confirmed CEMP stars belong to Group I ($A({\rm C}) \gtrsim 7.25$) and 97 to Group II ($A({\rm C}) \lesssim 7.25$) in the Yoon-Beers $A$(C)$-$[Fe/H] diagram. Moreover, we combine these data with Gaia EDR3 astrometric information to delineate new target-selection criteria, which have been applied to the Goodman/SOAR candidates, to more than double the efficiency for identification of bona-fide VMP and EMP stars in comparison to random draws from the B\&B catalog. We demonstrate that this target-selection approach can achieve success rates of $96\%$, $76\%$, $28\%$ and $4\%$ for [Fe/H] $\leq -1.5$, $\leq -2.0$, $\leq -2.5$ and $\leq -3.0$, respectively. Finally, we investigate the presence of dynamically interesting stars in our sample. We find that several VMP/EMP ([Fe/H] $\leq -2.5$) stars can be associated with either the disk system or halo substructures like Gaia-Sausage/Enceladus and Sequoia.


\end{abstract}



\section{Introduction}
\label{sec:intro}

\setcounter{footnote}{10}

Very metal-poor (VMP; [Fe/H]\footnote{Definition of elemental abundances for a star ($\star$) relative to the Sun ($\odot$): [A/B] $= \log (N_{\rm A}/N_{\rm B})_\star - \log (N_{\rm A}/N_{\rm B})_\odot$, where $N_{\rm A}$ ($N_{\rm B}$) is the number density of atoms of element A (B). The adopted composition of the Sun is from \citet{asplund2009}.} $< -2.0$) and extremely metal-poor (EMP; [Fe/H] $< -3.0$) stars are relics of the formation and evolution of the Galaxy, providing clues on the nucleosynthesis processes operating throughout its early history \citep{Beers2005}. Seminal efforts, focused towards discovering VMP and EMP stars in the past (e.g., the HK survey; \citealt{beers1985,beers1992}, and the Hamburg/ESO survey; \citealt{Christlieb2003,chris2008}), provided the majority of targets observed at high spectroscopic resolution over the last few decades (e.g., \citealt{norris1996, Hill2002,Cayrel2004, aoki2007, cohen2008,yong2013a,roederer2014}). Studies of these ancient stars have allowed stellar archaeologists to constrain the conditions for the chemical enrichment of the star-forming environments that existed in the nascent Milky Way \citep{frebel2015}.

\citet{beers1992} first noted the presence of a surprisingly large number of carbon-enhanced stars among their sample at the lowest metallicities.  As the sample sizes increased, it was recognized that the fraction of carbon-enhanced metal-poor (CEMP; [C/Fe] $> +0.7$ and [Fe/H] $< -1.0$) stars indeed rapidly increases with decreasing [Fe/H] \citep{norris1997, rossi1999, rossi2005,lucatello2006, lee2013, yong2013b,placco2014Carbon, yoon2018} and at greater distances from the Galactic plane ($|Z|_{\rm Gal}$; \citealt{frebel2006, carollo2012, lee2017, lee2019, yoon2018}). This behavior has been proposed to be related to the dual nature of the stellar halo (hereafter ``halo"; \citealt{carollo2007, carollo2010,beers2012}). It has been suggested that the majority of the CEMP stars in the [Fe/H] $\lesssim -2.5$ regime belong to the CEMP-no subclass, showing no enhancements in neutron-capture elements ([Ba/Fe] $< 0.0$; see \citealt{yoon2016, yoon2019}). These observations support the hypothesis that the CEMP-no stars are the direct descendants of massive Pop III stars that are now long vanished \citep{Ryan2005,aoki2007,Ito2013,Spite2013,Tominaga2014,keller2014,Frebel2015ApJ,Roederer2016CEMP,Placco2016A,Placco2016B,Aguado2018A,Ezzeddine2019}.

Another peculiarity found in the chemical-abundance profiles of some metal-poor stars is their enhancement in $r$-process (rapid neutron capture) elements (see \citealt{Sneden2008} and \citealt{frebel2018} for reviews on the topic). The qualitative aspects of the formation of these neutron-rich nuclei have been known for many decades (e.g., \citealt{burbidge1957,cameron1957}). However, the astrophysical site(s) in which the $r$-process occurs remained speculative up until the photometric and spectroscopic observations of the electromagnetic counterpart AT2017gfo \citep{Arcavi2017,drout2017, Pian2017,shappee2017,Smartt2017} of the gravitational wave event GW170817 \citep{Abbott2017A, Abbott2017B, Abbott2017C} of a neutron star merger. These authors concluded that this transient (kilonova) was powered by the radioactive decay of large amounts of $r$-process elements, in agreement with early theoretical predictions \citep{lattimer1974}. It has been suggested that neutron star mergers are the primary (perhaps the only) source of $r$-process enrichment in the Galaxy (e.g., \citealt{cote2018,Safarzadeh2019a, Banerjee2020,Dvorkin2020}), but other studies  \citep{Belczynski2018,JiandFrebel2018,Ji2019,Cote2019,Safarzadeh2019b,Haynes2019,Kobayashi2020} have provided evidence that additional sources may be involved.

In the Gaia era, proper motions (PMs) and parallaxes of exquisite quality have been made available for more than a billion stars \citep{TheGaiaMission}. The phase-space information, particularly from Gaia Data Release 2 (DR2; \citealt{gaiadr2}), combined with previously available high-resolution spectroscopic data, has enabled the exploration of the chemo-dynamical properties of these low-metallicity stars. It has been revealed that a large population of EMP and  ultra metal-poor (UMP; [Fe/H] $<-4.0$) stars are apparently kinematically connected (similar $|Z|_{\rm Gal}$ and rotational motion around the Galactic center) to the disk system \citep{sestito2019, sestito2020, Cordoni2020}, providing constraints on the assembly of the newborn Milky Way (redshift $z \gtrsim 2$) through its primordial building blocks \citep{DiMatteo2020,Sestito2021}.

Considering all of the above-mentioned advances, it is clear that numerous VMP and EMP stars are necessary to advance our understanding of the formation and evolution of the Galaxy. The pioneering objective-prism surveys \citep{bond1970, bond1980, bidelman1973, beers1985, beers1992, Christlieb2003, chris2008} have been responsible for the identification of thousands of VMP stars and several hundred EMP stars. Large spectroscopic surveys such as the Sloan Digital Sky Survey (SDSS; \citealt{sdssYork}) and its stellar-specific sub-survey Sloan Extension for Galactic Understanding and Exploration (SEGUE; \citealt{yanny2009}), the Large Sky Area Multi-object Fiber Spectroscopic Telescope (LAMOST; \citealt{LAMOST1,LAMOST3}; see also \citealt{lamostVMP}), and the Radial Velocity Experiment (RAVE; \citealt{rave}) have increased these numbers to tens of thousands of VMP and several thousand EMP stars. Ongoing and planned spectroscopic surveys in the near future (WEAVE: \citealt{WEAVE2012}; 4MOST: \citealt{4MOST2012, 4MOST2014}, Pristine: \citealt{Starkenburg2017PRISTINE, aguado2019, Youakim2020}; H3: \citealt{Conroy2019B, Conroy2019A}; SDSS-V/MWM: \citealt{Kollmeier2017}) are expected to expand these numbers further.

\begin{figure*}[pt!]
\centering
\includegraphics[scale=0.385]{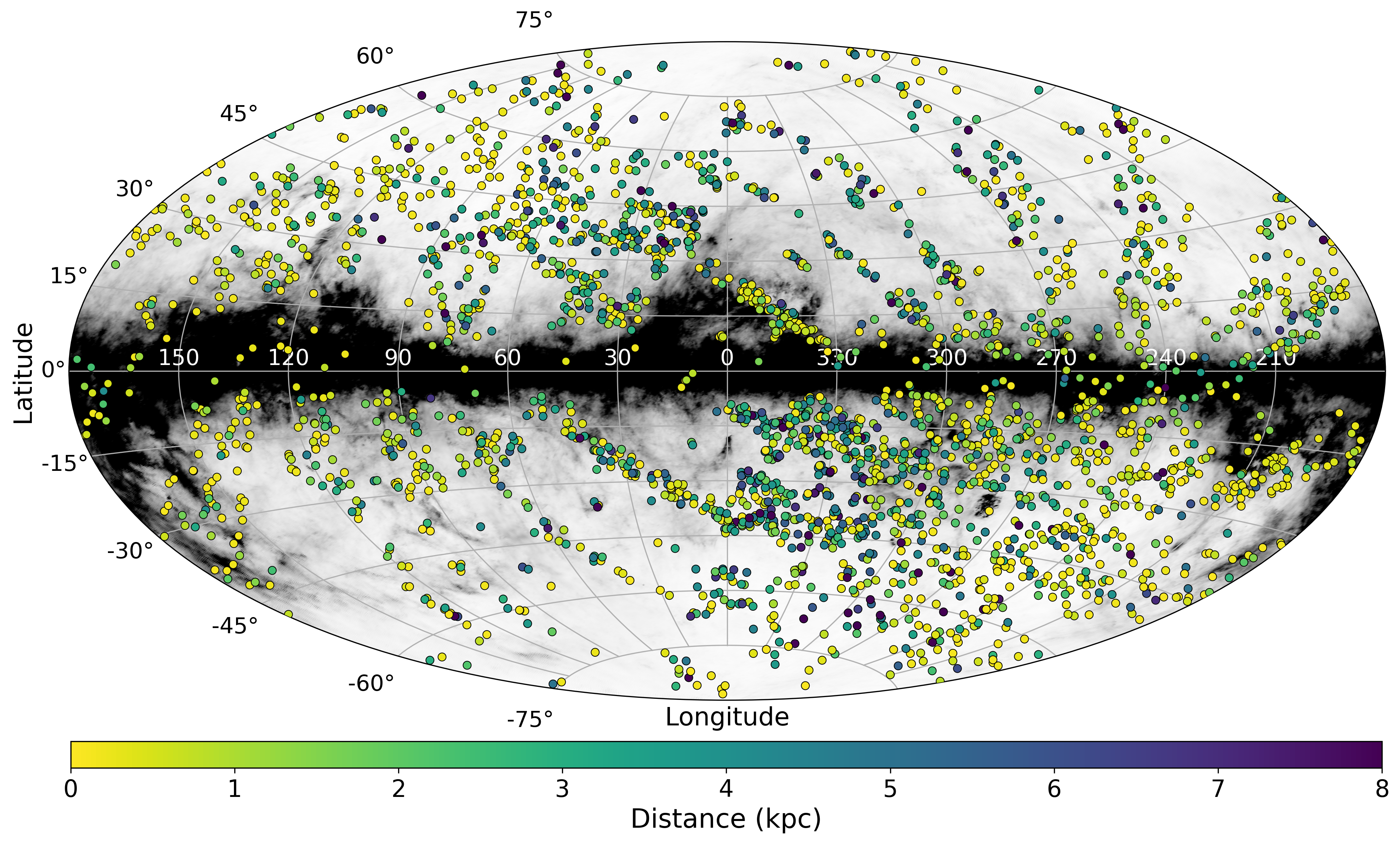}
\includegraphics[scale=0.345]{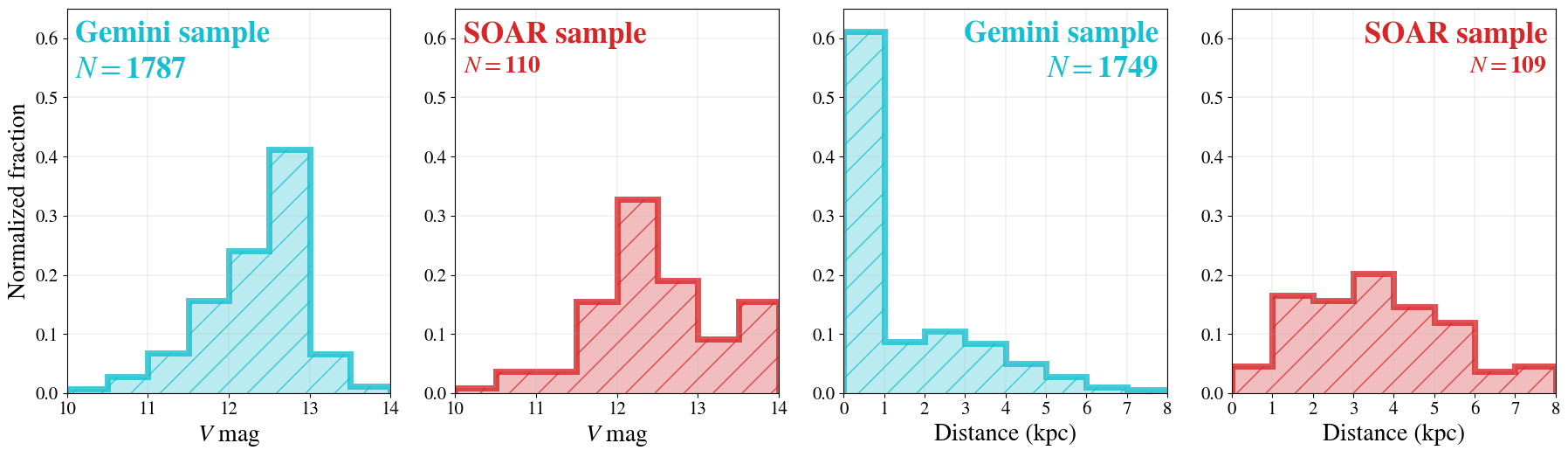}
\caption{{Top panel: Distribution of the Gemini (Section \ref{gemini}), SOAR (Section \ref{soar}), and \citetalias{placco2019} samples in the Galactic coordinate system, color-coded by heliocentric distances (Section \ref{sec:kinematics}). The background all-sky distribution of the Galactic reddening comes from the \citet{Schlegel1998} dust map, as re-calibrated by \citet{Schlafly2011}. The different gray scales represent $E(B-V)$ values from 0.0 (white) to 0.5 (black). Bottom panels: Distributions of $V$-band magnitudes (Section \ref{sec:obs}) and heliocentric distance estimates (Section \ref{sec:kinematics}) for the Gemini (cyan) and SOAR samples (red). The total number of stars represented in each histogram is also shown in their respective panels.}
\label{fig:projs}} 
\end{figure*}

More recently, the Best \& Brightest (B\&B; \citealt{beb2014}, and see also \citealt{Casey2015, Reggiani2020}) initiative has taken advantage of mid-infrared photometry from the Wide-field Infrared Survey Explorer (WISE; \citealt{wright2010}) mission, in combination with near-infrared photometry from the Two Micron All Sky Survey (2MASS; \citealt{2MASS}), to select almost 12000 low-metallicity candidates. From follow-up spectroscopy of $\sim$200 objects in this list, these authors obtained efficiencies of 33\% and 4\% in finding VMP and EMP stars, respectively. \citet[hereafter \citetalias{placco2019}]{placco2019} have incorporated magnitude (in the $V$-band), reddening ($E(B-V)$) and PM cuts in the target selection of their own follow-up of stars from the B\&B catalog. The more restrictive criteria yielded similar success rates, 42\% (VMP) and 2\% (EMP), for their much larger sample of $\sim$800 candidates, but including fainter targets. The VMP and EMP stars uncovered through the B\&B selection have served as targets for high-resolution spectroscopy conducted by the $R$-Process Alliance (e.g., \citealt{holmbeck2020} and references therein), as they are all brighter than $V = 14$ and can be readily observed with 2.5-m to 6.5-m class telescopes from the ground. 

With the advent of Gaia, we have the opportunity to combine astrometric (as discussed in \citealt{placco2018}, hereafter \citetalias{placco2018}) and photometric (WISE+2MASS/B\&B) information to increase the efficiency in identifying VMP and EMP stars. This should allow us to populate the carbon- and $r$-process-enhanced classes of metal-poor stars with bright objects much faster, enabling studies of their origins, constraining supernovae nucleosynthesis and chemical evolution models, and probing rare, chemically peculiar targets for future investigations. 

The goal of the present work is to identify VMP and EMP stars from the B\&B catalog, confirming (or not) their metal-poor nature via low-resolution ($R \sim 2000$) spectroscopy. We also seek to determine carbon and magnesium (representative of the $\alpha$ elements) abundances. Crucially, we incorporate Gaia Early Data Release 3 (EDR3; \citealt{GaiaEDR3Summary}) astrometry to investigate the effect of kinematic-based target selection on the success rates of finding low-metallicity stars. These criteria can be taken into account for ongoing and future searches for metal-poor stars in the Galaxy. The most interesting VMP and EMP stars vetted by this approach  will serve as targets for ongoing and forthcoming high-resolution spectroscopic campaigns. Finally, we revisit the behaviors of VMP star fractions, as functions of both $|Z|_{\rm Gal}$ and velocities, and the increase of CEMP star fractions with declining metallicity. We also investigate the dynamically interesting VMP/EMP stars with either disk- or halo-like orbits.  


\renewcommand{\arraystretch}{1.0}
\setlength{\tabcolsep}{0.47em}

\begin{table*}[ht!]
\centering
\caption{Coordinates and Observing Details  
}
\label{tab:coordinates}
\begin{tabular}{>{\footnotesize}c >{\footnotesize}c >{\footnotesize}c >{\footnotesize}r >{\footnotesize}c >{\footnotesize}r >{\footnotesize}c >{\footnotesize}c >{\footnotesize}c >{\footnotesize}c >{\footnotesize}c}
\hline
Star Name & Star Name & R.A. & \multicolumn{1}{c}{Decl.} & $l$ & \multicolumn{1}{c}{$b$} & Telescope & Instrument & Program ID    \\
(2MASS) & (Gaia EDR3) & (deg) & \multicolumn{1}{c}{(deg)} & (deg)    & \multicolumn{1}{c}{(deg)} & & & \\
\hline
\hline
00003305$-$7953389 & 4634573766005607552      & 0.1375       & $-$79.8942      & 305.7037 & $-$36.9587 & SOAR         & Goodman    & SO-2019B-013    \\
00020162$-$4430117 & 4994519032163925632      & 0.5067       & $-$44.5033      & 329.4221 & $-$69.9972 & SOAR         & Goodman    & SO-2019B-013    \\
00040305$-$6106367 & 4905632480654004608      & 1.0127       & $-$61.1102      & 312.9121 & $-$55.0903 & Gemini South & GMOS-S     & GS-2016A-Q-76   \\
00043646+4124062 & 384060304935385984       & 1.1519       & 41.4017       & 113.5728 & $-$20.6174 & Gemini North & GMOS-N     & GN-2017B-Q-75   \\
00045403+3524010 & 2876804519751163008      & 1.2251       & 35.4003       & 112.3464 & $-$26.5131 & Gemini North & GMOS-N     & GN-2016A-Q-75   \\
\hline
\end{tabular}
\begin{flushleft}
This table is available in its entirety in machine-readable form. 
\end{flushleft}
\end{table*}

\renewcommand{\arraystretch}{1.0}
\setlength{\tabcolsep}{0.6em}

\begin{table*}[ht!]
\centering
\caption{Colors, Magnitudes, and Reddening Estimates   
}
\label{tab:colors_magnitudes}

\begin{tabular}{>{\small}c >{\small}c >{\small}c >{\small}c >{\small}c >{\small}c >{\small}c >{\small}c >{\small}c }
\hline
Star Name             & Star Name                & $V$     & ($B-V$)      & $G$        & ($G_{\rm BP}-G_{\rm RP}$)        & $J$    & ($J-K$) & $E(B-V)$    \\
(2MASS) & (Gaia EDR3) & & & & & & &               \\
\hline
\hline
00003305$-$7953389 & 4634573766005607552      & 12.411 & 0.775  & 12.184   & 1.069    & 10.804 & 0.535 & 0.074    \\
00020162$-$4430117 & 4994519032163925632      & 12.631 & 0.807  & 12.397   & 1.067    & 11.007 & 0.567 & 0.011    \\
00040305$-$6106367 & 4905632480654004608      & 12.818 & 1.075  & 12.417   & 1.466    & 10.546 & 0.745 & 0.010    \\
00043646+4124062 & 384060304935385984       & 12.612 & 1.025  & 12.371   & 1.082    & 10.896 & 0.643 & 0.073    \\
00045403+3524010 & 2876804519751163008      & 12.174 & 0.698  & 11.925   & 1.011    & 10.576 & 0.513 & 0.063    \\
\hline
\end{tabular}
\begin{flushleft}
This table is available in its entirety in machine-readable form. 
\end{flushleft}
\end{table*}

This paper is outlined as follows. In Section \ref{sec:obs}, details of  target selection, observations, and data reduction are provided. Section \ref{sec:params} is dedicated to the estimation of stellar atmospheric parameters: effective temperature ($T_{\rm eff}$), surface gravity ($\log g$), and metallicity (as represented by [Fe/H]), and also elemental abundances of interest ([C/Fe] and [Mg/Fe]).  We explore the behaviors of these abundance ratios as functions of [Fe/H] in Section \ref{sec:abund}. In Section \ref{sec:kinematics}, we investigate the kinematics of the selected low-metallicity candidates and examine the improvements in the efficiency of finding VMP stars in the Galaxy. We also analyze the orbits of VMP/EMP in Section \ref{sec:kinematics} in the context of the recent literature. Finally, Section \ref{sec:conc} presents a summary of our conclusions.

\section{Target selection, observations, and data reduction}
\label{sec:obs}

All of our targets have been selected as metal-poor candidates by \citet{beb2014} as part of the B\&B Survey. A total of 1897 stars have been observed with either the Gemini Multi-Object Spectrographs (GMOS-N/S; \citealt{GMOSPaper1, GMOSPaper2}) or Goodman spectrograph \citep{clemens2004} at the Gemini (North/South; 8.1 m) and the Southern Astrophysical Research (SOAR; 4.1 m) telescopes, respectively. By design, all candidates are significantly bright ($V \lesssim 14.0$; bottom left panels of Figure \ref{fig:projs}), which makes them excellent for high-resolution follow-up. The typical signal-to-noise ratio (SNR) of our spectra is $\gtrsim 30$ per pixel at the wavelength region of the \ion{Ca}{2} K/H lines (3900$-$4000~\AA). Since the target-selection criteria were different for observations with each instrument, we divide our stars into a ``Gemini sample" (Section \ref{gemini}) and a ``SOAR sample" (Section \ref{soar}). The calibrations included bias frames, quartz flats, and arc-lamp exposures. The background subtraction, definition of aperture, extraction of the one-dimensional spectra, and wavelength calibrations for each spectrum have been conducted with standard \texttt{IRAF} 
\citep{IRAF1, IRAF2} packages.

Pertinent observational information for the stars in our samples are presented in Table \ref{tab:coordinates}, including 2MASS names and Gaia EDR3 \citep{GaiaEDR3Summary} IDs and coordinates. The different telescopes and instruments are also listed. Table \ref{tab:colors_magnitudes} includes relevant photometric information: $V$, $G$ and $J$ magnitudes and $B-V$, $G_{\rm BP}-G_{\rm RP}$ and $J-K$ colors from AAVSO Photometric All Sky Survey (APASS; \citealt{apass}) DR9, Gaia EDR3 \citep{GaiaEDR3Photometry}, and 2MASS, respectively. The reddening values (Table \ref{tab:colors_magnitudes}) have been estimated with the \citet{Schlegel1998} dust maps (top panel of Figure \ref{fig:projs}). These can be easily re-calibrated into other $E(B-V)$ systems (e.g., \citealt{Schlafly2011}).

\newpage

\begin{figure}[pt!]
\centering
\includegraphics[width=\columnwidth]{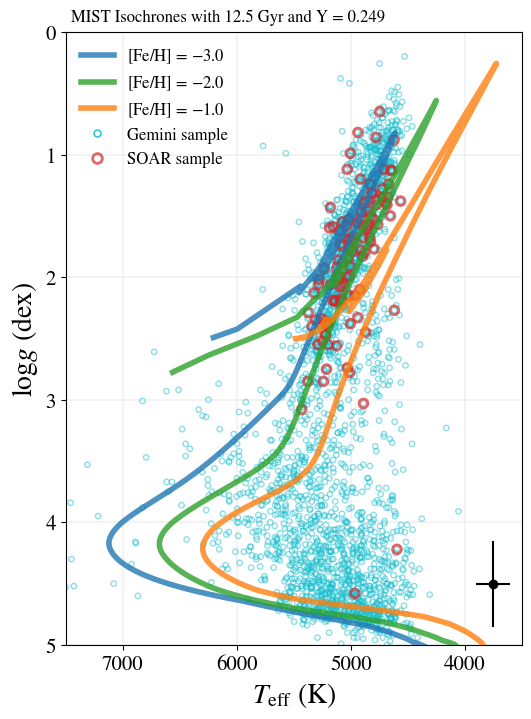}
\caption{{$T_{\rm eff}$ vs. $\log g$ diagram of the Gemini (cyan circles; Section \ref{gemini}) and SOAR (red symbols; Section \ref{soar}) samples. Typical uncertainties for these atmospheric parameters ($\pm$150 K for $T_{\rm eff}$ and $\pm$0.35 dex for $\log g$; Section \ref{sec:params}) are represented by the black dot with error bars in the bottom right corner. The colored lines are MESA Isochrones and Stellar Tracks (MIST; \citealt{MESA_0, MESA_1}) isochrones with varying metallicities. The orange, green, and blue isochrones represent $\rm [Fe/H] = -1.0$, $-2.0$, and $-3.0$, respectively. At the top, we list the rest of the conditions employed to generate the MIST stellar population models: $\rm Age = 12.5$ Gyr and $Y=0.249$ (primordial He abundance; \citealt{Planck2016}).} \label{fig:TeffLogg}} 
\end{figure}

\subsection{The Gemini Sample}
\label{gemini}

The majority (1787/1897; 94\%) of the spectra were acquired with the GMOS-N and GMOS-S spectrographs at Gemini North and South, respectively. The observations in this program were obtained between 2014A and 2019B; the various proposal IDs are listed in Table~\ref{tab:coordinates}. All observations used 0.5$''$ slits. During the semesters 2014A, 2015A, and 2015B, the spectra were obtained with the B1200 l mm$^{-1}$ G5301 (North) and G5321 (South) gratings, which led to a resolving power of $R \sim 2400$$-$$2800$. For all other observing runs, we have employed the B600 l mm$^{-1}$ G5307 (North) and G5323 (South) gratings, resulting in $R \sim 2000$$-$$2600$. The typical wavelength coverage of GMOS-N/S spectra is 3200$-$5800~\AA{}. We have been able to obtain stellar-parameter estimates for all but five stars in this sample (Section \ref{sec:params}). The cases for which we have not been able to estimate $T_{\rm eff}$ and $\log g$ were due to low-SNR ($\lesssim$10 per pixel at $\sim$4000 \AA) spectra and/or large mismatches between the color-based temperatures and the spectroscopic calibrations, which is expected for stars outside the $4000~\leq T_{\rm eff} \ (\rm K) \leq~7000$ range.

\subsection{The SOAR Sample}
\label{soar}

Unlike the Gemini sample, stars in the SOAR sample took into account phase-space information in the target selection. In \citetalias{placco2018}, PMs from Gaia DR1 \citep{gaiadr1} and line-of-sight\footnote{Throughout this work, we employ the ``line-of-sight" terminology instead of the more usual ``radial velocity" one. We reserve the latter for the radial component of the Galactocentric velocities in the cylindrical coordinate frame (see Section \ref{toomre}).} velocities ($V_{\rm los}$) from the RAVE DR5 \citep{Kunder2017} had already been used to explore this possibility. The success rate in finding low-metallicity ($\rm [Fe/H] \lesssim -1.5$) stars was higher for both larger $|Z|_{\rm Gal}$ and transverse velocities relative to the Sun ($V_{\rm T}$). Here, we develop this idea further, and propose an improved, more robust set of criteria. In order to take full advantage of both PMs and $V_{\rm los}$ from Gaia's past and future DRs, we introduce the quantity ``total available velocity" ($V_{\rm TAV}$), where:

\begin{align}
V_{\rm TAV} = \left\{ \begin{array}{lll} 
                \left(V_{\rm los}^2 + V_{\rm T}^2\right)^{1/2} \ {\rm if} \ V_{\rm los} \ {\rm and} \ V_{\rm T} \ {\rm are} \ {\rm available};  \\
                V_{\rm T} \ {\rm if} \ {\rm only} \ {\rm proper}  \ {\rm motions} \ {\rm are} \ {\rm available}; \\
                V_{\rm los} \ {\rm if} \ {\rm only} \ {\rm this} \ {\rm component} \ {\rm is} \ {\rm available}. \\
                \end{array} \right.
\end{align}

A detailed description, along with the advantages and limitations of the $V_{\rm TAV}$ parameter, is given in Section \ref{why}. From this definition, the candidates in the SOAR sample have been selected according to: $|Z|{_{\rm Gal}}~>~0.5$ kpc and $V_{\rm TAV} > 100$ km s$^{-1}$, derived from the Gaia DR2 data available at the time of observations. In Section \ref{sec:santucci}, we provide an in-depth exploration of the ${Z_{\rm Gal}}$ vs. $V_{\rm TAV}$ diagram, justifying these choices.


\renewcommand{\arraystretch}{1.0}
\setlength{\tabcolsep}{0.84em}

\begin{table*}[ht!]
\centering
\caption{Stellar Atmospheric Parameters and Abundances   
}
\label{tab:abunds}
\begin{tabular}{>{\normalsize}c >{\normalsize}c >{\normalsize}c >{\normalsize}r >{\normalsize}r >{\normalsize}r >{\normalsize}r >{\normalsize}c >{\normalsize}r }
\hline
Star Name & Star Name & $T_{\rm eff}$ & $\log g$  & [Fe/H] & [C/Fe] & [C/Fe]$_c$ & $A$(C)$_c$ & [Mg/Fe]  \\
(2MASS) & (Gaia EDR3) & (K) & (cgs) & & & & & \\
\hline
\hline
00003305$-$7953389 & 4634573766005607552& 5280    & 2.06 & $-$1.93& +0.13 & +0.17 & 6.67 & +0.38  \\
00020162$-$4430117 & 4994519032163925632& 5036    & 2.00 & $-$1.81& +0.30 & +0.37 & 6.99 & $\dots$  \\
00040305$-$6106367 & 4905632480654004608& 4648    & 4.13 & $-$1.27& +0.15 & +0.15 & 7.31 & $-$0.19 \\
00043646+4124062 & 384060304935385984 & 6068    & 3.51 & $-$1.29& +0.95 & +0.95 & 8.09 & +0.14  \\
00045403+3524010 & 2876804519751163008& 5455    & 3.52 & $-$2.59& +0.50 & +0.50 & 6.34 & +0.06  \\
\hline
\end{tabular}
\begin{flushleft}
[C/Fe]$_c$ and $A$(C)$_c$ values have been corrected for evolutionary status \citep{placco2014Carbon}. \\
This table is available in its entirety in machine-readable form. 
\end{flushleft}
\end{table*}

With Goodman/SOAR, 110 (6\% of the complete sample) metal-poor candidates were observed over the course of the 2018B and 2019B semesters (Table \ref{tab:coordinates}). The instrumental setup was similar to the one described in Section \ref{gemini}, including a 600 l mm$^{-1}$ grating and a 1.0$''$ long slit. The typical wavelength coverage for Goodman/SOAR spectra is 3600$-$6200~\AA{}, and the achieved resolution is $R\sim1300$. For the SOAR sample, stellar parameters have been derived for all but two stars.

\newpage

\section{Stellar Parameters and Abundances} 
\label{sec:params}

In order to determine the stellar atmospheric-parameter values ($T_{\rm eff}$, $\log g$, and [Fe/H]), we have followed the same approach as \citetalias{placco2018}, \citetalias{placco2019}, and \citet{Limberg2021}, where the fundamental proxy for metallicity is the \ion{Ca}{2} K line ($\sim$3933 \AA). We have employed the n-SSPP \citep{beers2014, beers2017}, a customized version of the Segue Stellar Parameter Pipeline (SSPP; \citealt{lee2008a, lee2008b, lee2011, lee2013}). This methodology consists of several routines (e.g., photometric calibrations and spectral-line indices) to derive estimates of the stellar parameters. It also compares the input spectra with a dense grid of synthetic ones in a $\chi^2$ minimization framework. The best set of values is then adopted taking into account the wavelength coverage of the analyzed spectrum, its SNR, and calculated uncertainties.

Considering both samples, the atmospheric parameters have been determined for all but seven stars (out of 1897; see Table \ref{tab:abunds}), as mentioned in Section \ref{sec:obs}. Based on empirical comparisons with high-resolution spectroscopic analyses \citep{placco2014High-Res, beers2014, beers2017}, the typical errors for $T_{\rm eff}$, $\log g$, and [Fe/H] are $\pm$150 K, $\pm$0.35 dex and $\pm$0.20 dex, respectively, for $\rm SNR \sim 30$ per pixel at $\sim$4000 \AA. These estimated atmospheric parameters can be visualized in Figure \ref{fig:TeffLogg}, where the $\log g$ vs. $T_{\rm eff}$ distribution is overlapped by MESA Isochrones and Stellar Tracks (MIST; \citealt{MESA_0, MESA_1}) models with varying metallicities. Overall, we have confirmed that 1064 ($56^{+2}_{-2}\%$\footnote{Uncertainties in the fractions are given by the \citet{Wilson1927} score approximation, which provides an estimate of the binomial proportion confidence intervals. This approximation is commonly used for small numbers statistics ($n \lesssim 40$), but is also comparable to other metrics when the analyzed sample is larger.}) of the newly observed stars are metal-poor ([Fe/H] $<-1.0$), 566 ($30^{+2}_{-2}\%$) are VMP, and 35 ($2^{+1}_{-1}\%$) are EMP. These are very similar proportions to what had been previously achieved by both \citet{beb2014} and \citetalias{placco2019} who performed spectroscopic follow-up with similar selection functions, targets from the B\&B Catalog. We reinforce that, although the targets observed at SOAR were chosen from kinematic criteria, these represent only a small portion of the complete (Gemini+SOAR) sample, hence these comparisons remain valid.

With the n-SSPP pipeline application, we have also obtained [C/Fe] estimates for most stars from both samples. The carbon-to-iron ratios are estimated from the strength of the CH $G$-band molecular feature ($\sim$4300~\AA). We have also calculated corrections for the [C/Fe] (and $A$(C)\footnote{Absolute abundance. $A$(X) $= \log\epsilon ({\rm X}) = \log (N_{\rm X}/N_{\rm H}) + 12$, where $N_{\rm X}$ and $N_{\rm H}$ are the number density of atoms of the given element and of hydrogen, respectively.}) following the prescriptions of \citet{placco2014Carbon}. These authors employed stellar evolution models \citep{Stancliffe2009} to account for the intrinsic carbon depletion (with consequent enhancement in nitrogen) in the atmospheres of metal-poor stars due to mixing with internal layers of material enriched by the CN cycle during the red giant branch phase \citep{Charbonnel1995}. The corrected carbon-abundance values are indicated by [C/Fe]$_c$ (or $A$(C)$_c$) throughout this work, and are listed in Table \ref{tab:abunds}. We also note that metallicities and carbon abundances of cooler ($T_{\rm eff} \lesssim 4500$ K) CEMP stars found in this work (and other low-resolution efforts) might be severely affected by the presence of ``carbon-veiling", which depresses the continuum in the wavelength region of the \ion{Ca}{2} K/H lines, hindering the metallicity determination (see discussion in \citealt{yoon2020}). The accuracy for [C/Fe] (and [C/Fe]$_c$) is 0.20 dex \citep{lee2013, beers2014, beers2017}. 

The n-SSPP derives the abundances of magnesium from the Mg I triplet absorption lines (located in the range of 5150$-$5200 \AA{}). For the Gemini sample, Mg-to-iron ratios have been obtained for 1104/1787 (62\%) of the stars. For the SOAR sample, 85/110 (77\%) stars had [$\alpha$/Fe] estimated. The cases for which these abundances are not determined are due to low-quality ($\rm SNR \lesssim 10$) spectra, or absorption features that are too weak to be distinguished from the underlying noise. The typical errors for [Mg/Fe] are also 0.20 dex (\citealt{lee2011}; \citetalias{placco2018, placco2019}). The [Mg/Fe] ratios are listed in Table \ref{tab:abunds}.

\begin{figure*}[!ht]
\centering
\includegraphics[scale=0.51]{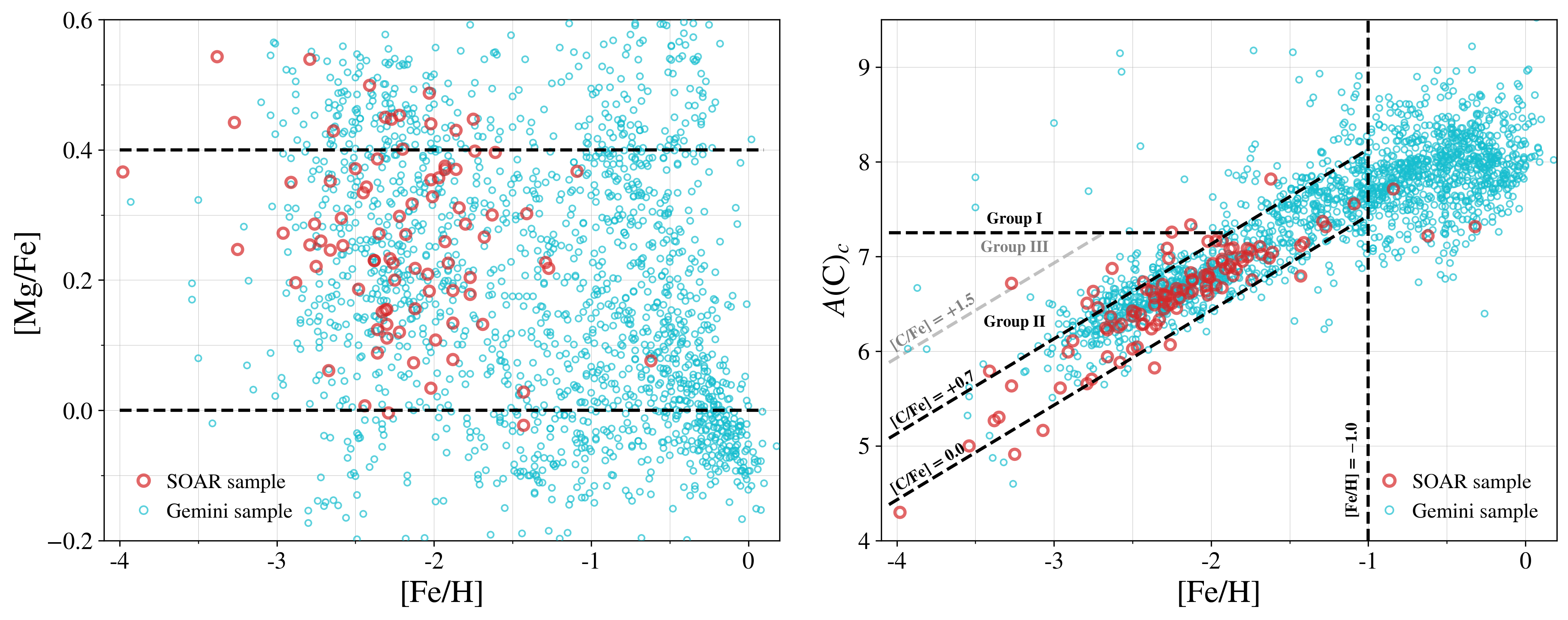}
\caption{{Left panel: Mg-to-iron ratio ([Mg/Fe]; Section \ref{sec:abund}) vs. metallicity ([Fe/H]; Section \ref{sec:params}). Dashed lines mark different levels of Mg enrichment. Cyan and red open circles correspond to stars in the Gemini (Section \ref{gemini}) and SOAR (Section \ref{soar}) samples, respectively, in a color scheme similar to Figures \ref{fig:projs} and \ref{fig:TeffLogg}. Right panel: Absolute carbon abundance ($A$(C)$_c$; Section \ref{sec:abund}), corrected for evolutionary effects \citep{placco2014Carbon}, vs. [Fe/H] diagram. Black dashed lines mark the classification for different levels of carbon enhancement according to \citet{yoon2016,yoon2019} and considered in this work; see text for details. For convenience, we mark the characteristic region of Group III with a gray line. The vertical line marks the metal-poor limit at $\rm[Fe/H] = -1.0$, for reference.}
\label{fig:abunds}} 
\end{figure*}

\section{Carbon and \texorpdfstring{$\alpha-$} eelements}
\label{sec:abund}

Carbon-to-iron ratios in low-metallicity stars can constrain their different formation scenarios (see, e.g., \citealt{Aoki2002, aoki2007, Nomoto2013, Hansen2016B}). The $\alpha$-element abundances are crucial for our classical understanding of Galactic chemical evolution (e.g., \citealt{Chiappini1997}). Furthermore, the Mg-to-carbon ([Mg/C]) ratio serves as a diagnostic for the so-called ``mono-enriched" population ([Fe/H] $\lesssim -2.5$ and [Mg/C] $\lesssim -1.0$; \citealt{hartwig2018}, see also \citealt{Rasmussen2020}) which are, potentially, genuine second-generation stars. Figure \ref{fig:abunds} provides plots of the [Mg/Fe] and $A$(C)$_c$ distributions as functions of [Fe/H] (left and right panels, respectively). 

We first discuss the $\alpha$-element abundances. Within the wavelength coverage and resolution of our spectra, the estimated [Mg/Fe] is the main representative of the $\alpha$ process. In the metal-rich regime, there is an accumulation of stars at [Mg/Fe] $\sim 0.0$ and [Fe/H] $\gtrsim -0.5$, associated with the (low-$\alpha$) thin disk. We also note the presence of stars related to the (high-$\alpha$) thick disk at [Mg/Fe] $\sim +0.3$ and [Fe/H] $\gtrsim -1.0$. Since we focus on metal-poor stars, further exploration of these components is beyond the scope of the present work. All estimated magnesium-to-iron ratios are in the range $-0.2 <$ [Mg/Fe] $< +0.6$, which is expected from Galactic chemical-evolution models based on high-precision spectroscopy of $\alpha$ elements \citep{Zhao2016, Reggiani2017}.

Regarding the carbon abundances, we have explored the Yoon-Beers $A$(C) vs. [Fe/H] diagram following the suggested classification of \citet{yoon2016, yoon2019}. For consistency, we have applied the criteria outlined by \citetalias{placco2018}, also employed in \citetalias{placco2019}, presented in Figure \ref{fig:abunds} (dashed lines). Besides the $\rm[Fe/H] <-1.0$ and ${\rm[C/Fe]}_c >+0.7$ conditions, we consider CEMP stars to be members of the so-called Group I if $A$(C)$_c$ $\gtrsim 7.25$. Since the carbon enrichment of these stars' atmospheres are usually accompanied by enhancement in their abundances of slow neutron-capture ($s$-process) elements \citep{Spite2013, Bonifacio2015,hansen2015,Cruz2018}, they are generally associated with the CEMP-$s$ ([Ba/Fe] $>+1.0$ and [Ba/Eu] $>+0.5$; \citealt{Beers2005}) subclass of CEMP stars. Such a chemical profile is thought to be the result of mass transfer from a binary companion in the asymptotic giant branch (AGB) phase \citep{Suda2004,Ryan2005,Lucatello2005, Bisterzo2011,Allen2012,Placco2013, Starkenburg2014, Hansen2016B}. On the other hand, CEMP stars with $A$(C)$_c$ $\lesssim 7.25$ are classified as Group II. Unlike the CEMP-$s$ category, these low-$A$(C) objects do not present over-abundances of $s$-process elements ([Ba/Fe] $<0.0$; \citealt{Beers2005}), and are known as CEMP-no stars. This abundance pattern is the result of pollution of these stars' birth environments through a single EMP (perhaps metal-free; \citealt{Heger2010}) core-collapse supernovae episode (see \citealt{Nomoto2013} for a review). Typically, the $A$(C) vs. [Fe/H] diagram is further partitioned with a Group III, analogous to Group II, but with [C/Fe]$_c$ $\gtrsim +1.5$ \citep{yoon2016,yoon2019}. These same authors argued that CEMP-$s$ and CEMP-no stars could be distinguished from $A$(C) alone with $\sim$90\% purity. We also note that the Group III stars are generally found at the lowest metallicities ([Fe/H] $\lesssim -3.0$), for which we have only a handful of stars in our samples (Section \ref{sec:params}).  Hence, for simplicity, we have not considered the Group III classification in this work. 

We have confirmed that a total of 191 stars in the Gemini$+$SOAR sample are CEMP  stars\footnote{Note that CEMP stars can be further divided into the CEMP-$r$ ([Eu/Fe] $>+0.3$ and [Ba/Eu] $<0.0$) and CEMP-$r/s$ or CEMP-$i$ ($0.0 <$ [Ba/Eu] $<+0.5$) categories \citep{hansen2018, ezzedine2020}. However, these are not considered in this work given the impossibility of classifying them without knowledge of their Ba and Eu abundances.}. Out of these, 94 belong to the (Group I) CEMP-$s$ subclass according to the criteria described above. The remaining 97 of our CEMP stars can be classified as Group II (CEMP-no). From Figure \ref{fig:abunds}, not only the overall fraction of CEMP stars increases with decreasing metallicity, but also the CEMP-no class dominates the [Fe/H] $\lesssim -2.3$ regime, similar to \citet{yoon2018}, which is in line with the expectation that the emergence of CEMP-$s$ stars in the Galaxy is driven by AGB evolution \citep{Herwig2005} timescales. On the other hand, all CEMP stars above [Fe/H] $\gtrsim -1.8$ are CEMP-$s$ (Group I).

Finally, we have confirmed the trend of increasing CEMP star fractions as a function of declining metallicity. For [Fe/H] $<-2.5$, we have found a CEMP star fraction of $32^{+6}_{-6}\%$, compatible ($1\sigma$) with both \citetalias{placco2019} and \citet{beb2014}. We have also found excellent agreement between our results, those from \citetalias{placco2019}, and the high-resolution effort of \citet{placco2014Carbon}. For [Fe/H]~$<-2.0$, we have obtained $19^{+3}_{-3}\%$. In the [Fe/H]~$<-3.0$ regime, for which we have very few stars, we have calculated $43^{+16}_{-15}\%$.  

\section{Kinematics of the Metal-poor Candidates}
\label{sec:kinematics}

Astrometry can be used to improve the efficiency in the search for VMP stars in the Galaxy. In \citetalias{placco2018}, the Tycho-Gaia Astrometric Solution database \citep{Lindegren2016} was combined with distances from \citet{Astraatmadja2016} to calculate the Cartesian Galactic positions ($X_{\rm Gal}$, $Y_{\rm Gal}$, and $Z_{\rm Gal}$) and velocities (particularly $V_{\rm T}$). For $V_{\rm T} \geq 75$ km s$^{-1}$, the authors noticed a considerable increase in the fractions of metal-poor and VMP stars in their own validation sample of low-metallicity stars from RAVE DR5. However, the full potential of this approach was still not clear at the time, because ($i$) those targets were selected with prior information of their metallicities from RAVE's moderate-resolution ($R \sim 7500$) spectroscopy and ($ii$) the limited coverage and quality of Gaia DR1 data did not allow for a confident (with small errors) investigation of the majority of their studied stars. In \citetalias{placco2019}, despite this knowledge, the better astrometry from Gaia DR2 was still not available at the time of observations, hence only a simple PM (plus $V$-band magnitude and reddening) cut was applied. In this work, we seek to explore more fully the influence of kinematic-based target selection in finding VMP and EMP stars, taking advantage of Gaia EDR3 PMs and $V_{\rm los}$. The results of these analyses will be (and are being) employed in future (and ongoing) observational campaigns. 

We have cross-matched both the Gemini and SOAR samples with Gaia EDR3 to acquire accurate $V_{\rm los}$, parallaxes, and PMs (Table \ref{tab:kinematics}). These measured parameters have been converted to the Cartesian Galactic phase-space positions and velocities using the \texttt{Astropy} package \citep{astropy, astropy2018}. The assumed in-plane distance from the Sun to the Galactic center is $R_{\odot}=8.2$ kpc \citep{BlandHawthorn2016}, which is compatible with \citet{GRAVITY2019, GRAVITY2020}. We recall that stars in the SOAR sample have been observed with prior knowledge of their kinematics, derived from Gaia DR2 data. For the Gemini sample, matches with Gaia EDR3 have been found for all but two stars (Section \ref{sec:obs}). Out of these, 1766/1787 (99\%) have non-negative parallaxes. The distances have been calculated through the inversion of these parallaxes\footnote{During the reviewing process of this paper, distances derived from a probabilistic approach have been made available by \citet{BailerJones2021} for sources in Gaia EDR3. Since our sample is fairly local ($\sim$80\% within 4 kpc; Figure \ref{fig:projs}), the differences between these authors' calculated distances and those adopted here are negligible ($>$10\% for only a couple of stars). In fact, even considering the full sample (up to $\sim$8 kpc), only $\sim$5\% of the stars with \texttt{parallax\_over\_error} $\geq 5$ show differences $>$10\% in their distance estimates.} after re-calibration ($+0.017$ mas\footnote{The parallax offset depends in a non-linear, non-trivial way on the ($G$-band) magnitude and color ($G_{\rm BP}-G_{\rm RP}$) of the source. For simplicity, we adopt a global correction equal to the median of quasar's parallaxes in Gaia EDR3 \citep{Lindegren2020b}.}; \citealt{Lindegren2020b}). For the purpose of our investigation, we have applied a relative error cut, keeping stars with \texttt{parallax\_over\_error} $\geq~1$ (see \citealt{Lindegren2020a}). Despite being very permissive in comparison to the recent literature, this uncertainty cut is consistent with \citetalias{placco2018}, so our study can be directly comparable to theirs, and has the objective of not removing too many candidates. However, we note that, thanks to the improved astrometry of Gaia EDR3 in comparison to DR2, only a handful ($\sim$80) stars in the full (Gemini+SOAR+\citetalias{placco2019}) sample have \texttt{parallax\_over\_error} $\leq~5$ (or missing/negative values). Furthermore, $\sim$91\% of all stars observed have \texttt{parallax\_over\_error} $\geq~10$, providing reliability to our kinematic/dynamical calculations. \\

\subsection[]{Why $V_{\rm TAV}$?}
\label{why}

\begin{figure}[htp!]
\centering
\includegraphics[width=\columnwidth]{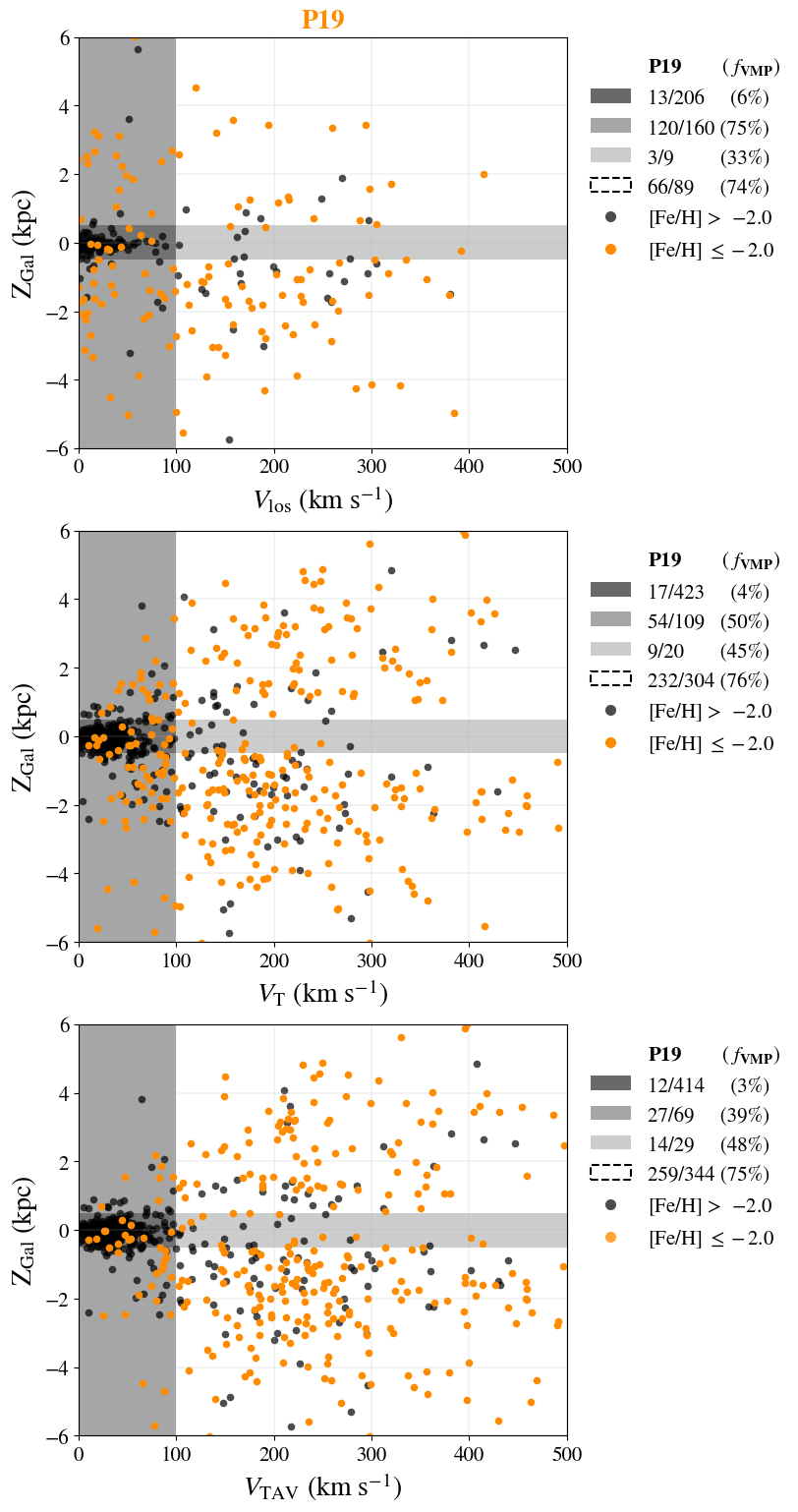}
\caption{{Diagrams of the \citetalias{placco2019} sample for $Z_{\rm Gal}$ vs. different components of velocity: $V_{\rm los}$ (top panel), $V_{\rm T}$ (middle), and $V_{\rm TAV}$ (bottom panel). To the right of each panel, the fractions of VMP stars for each gray-shaded and white regions are presented. The black and orange dots represent stars with [Fe/H] $> -2.0$ and VMP stars, respectively
}
\label{fig:santucci_EXAMPLE}} 
\end{figure}

In this work, we build on the results presented by \citetalias{placco2018}. One of the refinements that we have made is the introduction of the $V_{\rm TAV}$ quantity (Section \ref{soar}), which demands either PMs or $V_{\rm los}$, but makes use of their combination whenever both are available. With the goal of performing a comprehensive analysis, we have divided the $Z_{\rm Gal}$ vs. velocity diagrams into four different regions, represented with distinct grayish and white colors in Figure \ref{fig:santucci_EXAMPLE}. We find that high efficiencies of VMP and EMP detections are achieved even if only one velocity component is known. The $\pm$0.5 kpc stripe at the center of all panels is motivated by the characteristic scale height of thin-disk stars around the Galactic plane \citep{Recio-Blanco2014, LiZhao2018}, which is also consistent with the analysis of \citetalias{placco2018}. The separation at $100$ km s$^{-1}$ approximately marks the end of the metal-rich-dominated ([Fe/H] $> -2.0$) portion of the diagrams and can be considered a velocity limit between thin-disk ($\lesssim$100 km s$^{-1}$) and thick disk/halo populations, which is visible from the large concentrations of black dots in Figure \ref{fig:santucci_EXAMPLE}.

Indeed, it is expected that higher values of velocity should yield greater fractions of genuine halo stars (see \citealt{koppelman2018} and \citealt{Posti2018} for recent discussions). However, as is made clear below, the chosen boundaries already produce quite high success rates.
The combination of these cuts at $\pm$0.5 kpc and 100 km s$^{-1}$ defines the various areas in Figure \ref{fig:santucci_EXAMPLE}. Then, we calculate the fractions of stars showing [Fe/H] $\leq -2.0$ ($f_{\rm VMP}$) in each of these regions. We take these fractions as representative of the efficiency of finding VMP stars in each different portion of the $Z_{\rm Gal}$ vs. velocity diagrams for stars originally identified as possibly metal-poor on the basis of the B\&B photometric selection.

In the top panel of Figure \ref{fig:santucci_EXAMPLE}, only $V_{\rm los}$ is employed, in combination with $Z_{\rm Gal}$, to explore the $f_{\rm VMP}$ in the \citetalias{placco2019} sample. Although high success rates already appear for $|Z|_{\rm Gal} \geq 0.5$ kpc, the number of available candidates in the \citetalias{placco2019} sample is reduced to approximately half of its total. In the middle panel of Figure \ref{fig:santucci_EXAMPLE}, the ($Z_{\rm Gal}$, $V_{\rm T}$) space is displayed, similar to the figures presented by \citetalias{placco2018}. We note that almost the entire \citetalias{placco2019} sample has suitable PMs and parallaxes from Gaia EDR3, allowing us to calculate accurate $V_{\rm T}$ for these stars. The $f_{\rm VMP}$ in the white regions of the ($Z_{\rm Gal}$, $V_{\rm T}$) diagram remains statistically the same as the ($Z_{\rm Gal}$, $V_{\rm los}$) one, but the number of stars has more than tripled, which is advantageous for planning an observational campaign. 

The bottom panel of Figure \ref{fig:santucci_EXAMPLE} shows the $Z_{\rm Gal}$ vs. $V_{\rm TAV}$ diagram for the \citetalias{placco2019} sample. Once more, the $f_{\rm VMP}$ in each area of the diagram is similar to previous panels, but there is an increase of $\sim$13\% in the number of stars in the $|Z|_{\rm Gal} \geq 0.5$ kpc and $V_{\rm TAV} \geq 100$ km s$^{-1}$ region, while maintaining high efficiency, almost double the $f_{\rm VMP}$ achieved by \citetalias{placco2019}. This finding inspired the target-selection criteria for observations with Goodman/SOAR. Even though we had the prospect of improvements, we were surprised by these results, since the B\&B catalog is by-design biased towards metal-poor stars. In general, from top to bottom, one can notice the increasing number of points in each diagram of Figure \ref{fig:santucci_EXAMPLE}, but maintaining statistically equivalent success rates for the different regions, demonstrating the advantages of employing the $V_{\rm TAV}$ parameter.


\renewcommand{\arraystretch}{1.0}
\setlength{\tabcolsep}{0.66em}

\begin{table*}[ht!]
\centering
\caption{Phase-space Information from Gaia EDR3  
}
\label{tab:kinematics}
\begin{tabular}{>{\small}c >{\small}c >{\small}r >{\small}c >{\small}c >{\small}r >{\small}r >{\small}r >{\small}r}
\hline 
Star Name & Star Name & \multicolumn{1}{c}{$V_{\rm los}$} & $d_{\rm helio}$  & $\sigma_d$ & \multicolumn{1}{c}{PM$_{\rm R.A.}$} & \multicolumn{1}{c}{PM$_{\rm Decl.}$} & \multicolumn{1}{c}{$Z_{\rm Gal}$} & \multicolumn{1}{c}{$V_{\rm TAV}$}    \\
(2MASS) & (Gaia EDR3) & \multicolumn{1}{c}{(km s$^{-1}$)} & (kpc) & (kpc) & \multicolumn{1}{c}{(mas yr$^{-1}$)} & \multicolumn{1}{c}{(mas yr$^{-1}$)} & \multicolumn{1}{c}{(kpc)} & \multicolumn{1}{c}{(km s$^{-1}$)} \\
\hline
\hline
00003305$-$7953389 & 4634573766005607552      & 269.3     & 2.51      & 0.06      & $-$4.538    & $-$4.538    & $-$1.48     & 274.3      \\
00020162$-$4430117 & 4994519032163925632      & 58.5      & 3.13      & 0.19      & 14.656    & 14.656    & $-$2.92     & 202.6      \\
00040305$-$6106367 & 4905632480654004608      & $-$6.0      & 0.12      & 0.00      & 53.448    & 53.448    & $-$0.07     & 15.7       \\
00043646+4124062 & 384060304935385984       & $-$13.7     & 1.01      & 0.01      & $-$7.058    & $-$7.058    & $-$0.33     & 28.8       \\
00045403+3524010 & 2876804519751163008      & $-$91.5     & 1.06      & 0.02      & 9.525     & 9.525     & $-$0.45     & 154.2      \\
\hline
\end{tabular}
\begin{flushleft}
This table is available in its entirety in machine-readable form. 
\end{flushleft}
\end{table*}

\begin{figure*}[!ht]
\centering 
\includegraphics[scale=0.40]{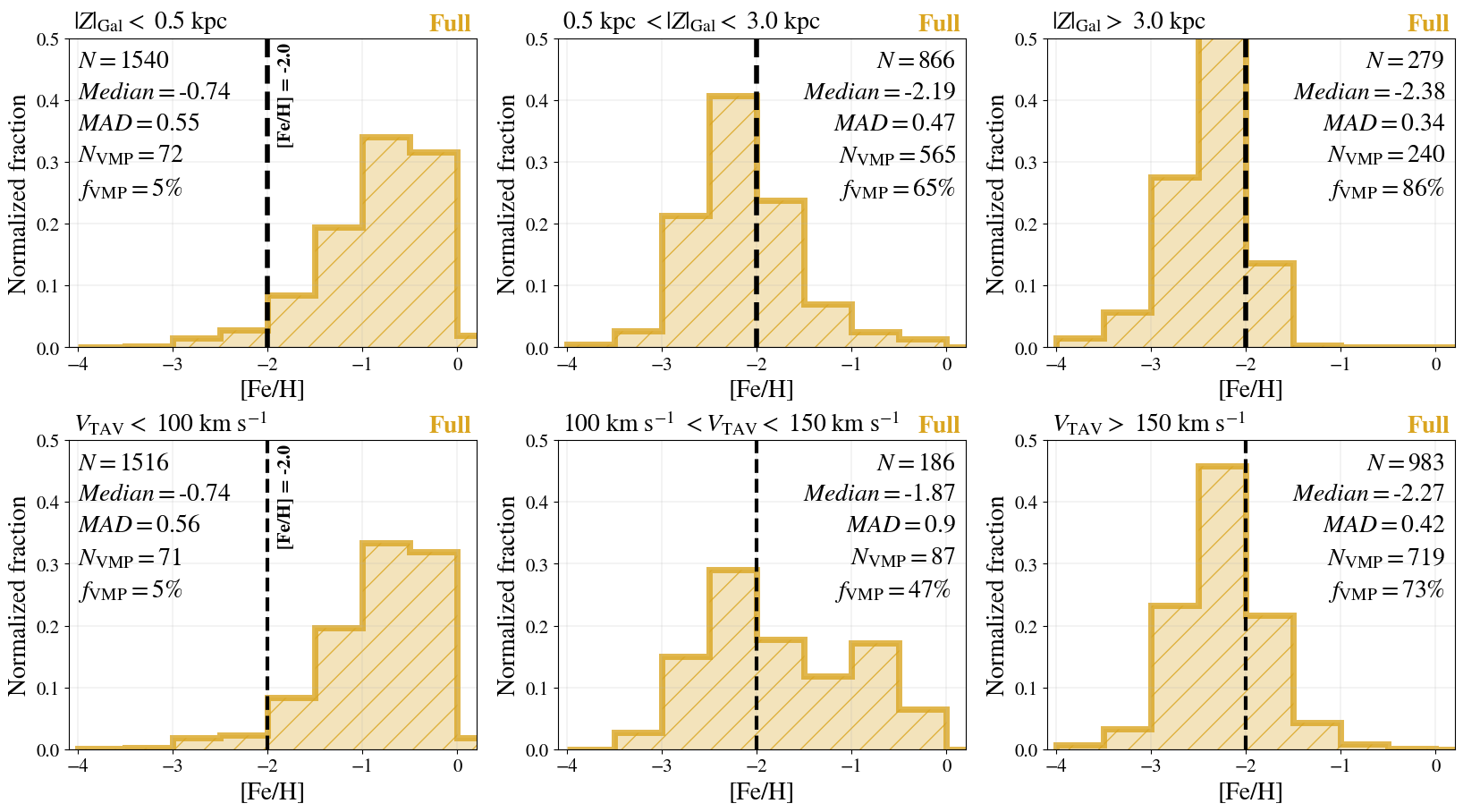}
\caption{{Metallicity distribution functions (MDFs) of stars observed from the B\&B catalog. The histograms present the normalized fractions; the total number of stars in each panel ($N$) is provided. Top panels: MDFs of the full (Gemini$+$SOAR$+$\citetalias{placco2019}) sample for the slices: $|Z|_{\rm Gal} < 0.5$ kpc, $0.5 < |Z|_{\rm Gal} \ \rm (kpc) < 3.0$, and $|Z|_{\rm Gal} > 3.0$ kpc (left, middle, and right, respectively). Bottom panels: MDFs of the full sample, but for various ranges of $V_{\rm TAV}$: $V_{\rm TAV} < 100$ km s$^{-1}$, $100 < V_{\rm TAV} \ \rm (km \ s^{-1}) < 150$ and $V_{\rm TAV} > 150$ km s$^{-1}$ (left, middle, and right, respectively). The median [Fe/H], median absolute deviation (MAD), the total number ($N_{\rm VMP}$), and the fraction of VMP stars ($f_{\rm VMP}$) are shown in each panel. The dashed vertical lines mark the VMP limit at [Fe/H] = $-2.0$, for reference. }
\label{fig:hists}} 
\end{figure*}

\subsection[]{$Z_{\rm Gal}$ vs. $V_{\rm TAV}$ Diagrams and Final Success Rates}
\label{sec:santucci}

The individual impact of variations in $|Z|_{\rm Gal}$ and $V_{\rm TAV}$ in the metallicity distribution function (MDF) of the full B\&B sample are presented from the top and bottom rows of Figure \ref{fig:hists}, respectively. In the thin-disk-like regions ($|Z|_{\rm Gal} < 0.5$ kpc and $V_{\rm TAV} < 100$ km s$^{-1}$; left column), the MDFs peak at [Fe/H] $\sim -0.7$ and the $f_{\rm VMP}$ is only $\sim$5\%. However, as we move to higher values of $|Z|_{\rm Gal}$, the peak of the MDF is shifted to the VMP regime already at $|Z|_{\rm Gal} > 0.5$ kpc (middle). The $f_{\rm VMP}$ reaches $86\%$ for $|Z|_{\rm Gal} > 3.0$ kpc (right), which reflects the lower metallicity of the Galactic halo in comparison to the canonical disk system, despite the already low-metallicity-biased B\&B selection. For the $V_{\rm TAV}$ quantity, this transition is smoother, noticeable from the extended MDF within the range $100 < V_{\rm TAV} \ \rm (km \ s^{-1})<150$ (middle column). However, the $f_{\rm VMP}$ significantly increases at $V_{\rm TAV} > 150$~km~s$^{-1}$ (right). 

\begin{figure*}[!ht]
\centering
\includegraphics[scale=0.48]{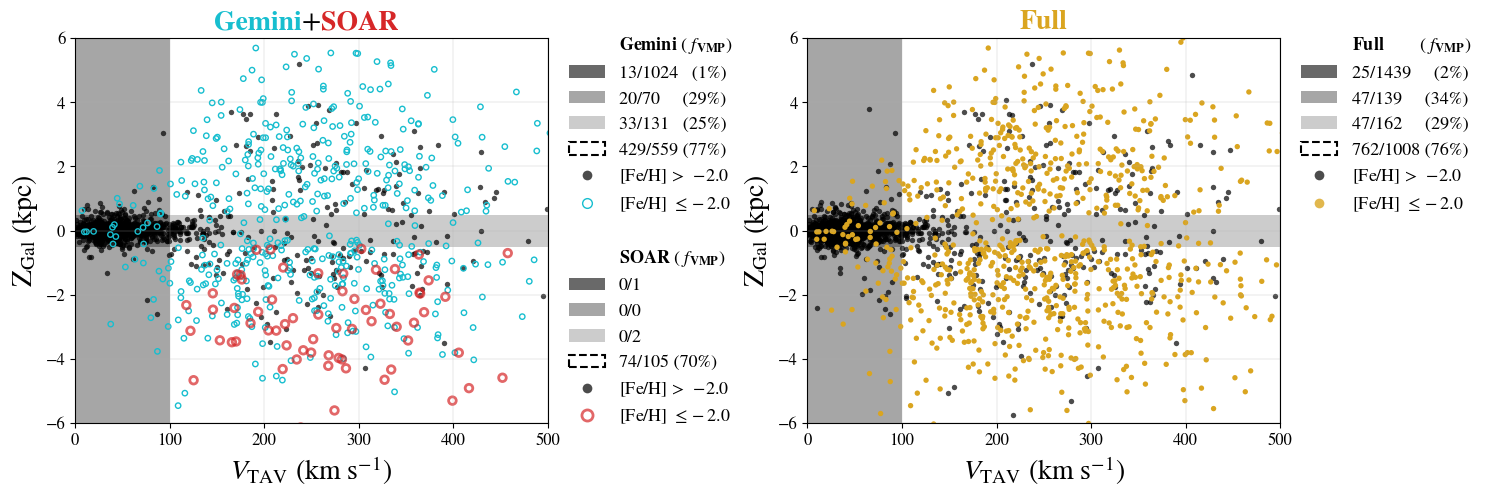}
\caption{{$Z_{\rm Gal}$ vs. $V_{\rm TAV}$ diagrams. Left: Distribution of stars from Gemini (Section \ref{gemini}) and SOAR (Section \ref{soar}) samples. Right: Distribution of stars from the full (Gemini+SOAR+\citetalias{placco2019}) sample. Colored symbols are VMP stars, while the black ones have [Fe/H]~$>-2.0$. Both panels are shown with a color scheme similar to Figures \ref{fig:projs}, \ref{fig:TeffLogg}, \ref{fig:abunds}, and \ref{fig:hists}. Specific VMP fractions are also provided for each gray-shaded and white regions (see text and Table \ref{tab:regions} for details) to the right of each panel. }
\label{fig:santucci}} 
\end{figure*}


\renewcommand{\arraystretch}{1.0}
\setlength{\tabcolsep}{2.0em}

\begin{table*}[ht!]
\centering
\caption{Regions of the $Z_{\rm Gal}$ vs. $V_{\rm TAV}$ Diagrams and Final Success Rates for Different Metallicities 
}
\label{tab:regions}
\begin{tabular}{>{\small}c >{\small}c >{\small}c >{\small}c >{\small}c >{\small}c >{\small}c }
\hline  
Region & $|Z|_{\rm Gal}$ & \multicolumn{1}{c}{$V_{\rm TAV}$} & [Fe/H]  & [Fe/H] & [Fe/H] & [Fe/H]     \\
(Color) & (kpc) & \multicolumn{1}{c}{(km s$^{-1}$)} & $\leq -1.5$ & $\leq -2.0$ & $\leq -2.5$ & $\leq -3.0$ \\
\hline
\hline
Dark gray   & $< 0.5$    & $< 100$    & $8^{+2}_{-1}\%$ & $2^{+1}_{-1}\%$ & $1^{+1}_{-0}\%$ & $0^{+0}_{-0}\%$ \\
Medium gray & $\geq 0.5$ & $< 100$    & $58^{+8}_{-8}\%$ & $34^{+8}_{-7}\%$ & $17^{+7}_{-5}\%$ & $3^{+4}_{-2}\%$ \\
Light gray  & $< 0.5$    & $\geq 100$ & $52^{+8}_{-8}\%$ & $29^{+7}_{-6}\%$ & $9^{+6}_{-4}\%$ & $1^{+3}_{-1}\%$ \\
White       & $\geq 0.5$ & $\geq 100$ & $96^{+1}_{-1}\%$ & $76^{+3}_{-3}\%$ & $28^{+3}_{-3}\%$ & $4^{+1}_{-1}\%$ \\ 
\hline
\end{tabular}
\begin{flushleft}
\end{flushleft}
\end{table*}

Figure \ref{fig:santucci} presents the $Z_{\rm Gal}$ vs. $V_{\rm TAV}$ diagrams for the Gemini and SOAR samples. Colored symbols are VMP stars spectroscopically confirmed by the present work, following the same color scheme as Figures \ref{fig:projs}, \ref{fig:TeffLogg}, \ref{fig:abunds}, and \ref{fig:hists}, while the black dots are those with [Fe/H] $>-2.0$. The left panel shows the distributions of stars from Gemini and SOAR samples in this parameter space. Given the target selection described in Section \ref{soar}, the majority of stars observed from SOAR occupy the most interesting regions of the $Z_{\rm Gal}$ vs. $V_{\rm TAV}$ diagram. Interestingly, the few cases in which they are outside the VMP regions originate from differences between the Gaia DR2 and EDR3 parameters. The right panel presents the ($Z_{\rm Gal}$, $V_{\rm TAV}$) space of the full sample, including that from \citetalias{placco2019}.

The $f_{\rm VMP}$ in each of the gray-shaded and white regions of the $Z_{\rm Gal}$ vs. $V_{\rm TAV}$ diagrams are also provided in Figure \ref{fig:santucci}. Table \ref{tab:regions} lists the final success rates for each region and for various metallicity regimes ([Fe/H] $\leq-1.5$, $\leq-2.0$, $\leq-2.5$, and $\leq-3.0$), along with their respective $|Z|_{\rm Gal}$ and $V_{\rm TAV}$ boundaries. These final efficiencies are drawn from the full sample for better statistics. Particularly in the white regions ($|Z|_{\rm Gal} \geq 0.5$ kpc and $V_{\rm TAV}~\geq~100$~km~s$^{-1}$), the thick disk/halo-like ones, the fractions of stars within [Fe/H] $\leq-1.5$, $\leq-2.0$, $\leq-2.5$, and $\leq-3.0$ are $96\%$, $76\%$, $28\%$, and $4\%$, respectively. Such effectiveness in the search for low-metallicity stars in the Galaxy are only rivaled by those with prior narrow-band photometry at the wavelength of the \ion{Ca}{2} K line (e.g., \citealt{Youakim2017, aguado2019, DaCosta2019}).
In the near future, we expect to  be able to combine the astrometric information from Gaia's DRs with narrow-band photometry from, e.g., S-PLUS (Southern-Photometric Local Universe Survey; \citealt{S-PLUS2019}), J-PLUS (Javalambre-Photometric Local Universe Survey; \citealt{J-PLUS2019}), and J-PAS (Javalambre-Physics of the Accelerating Universe Astrophysical Survey; \citealt{JPAS2020}) to obtain even better success rates, particularly in the EMP regime.

\begin{figure*}[htp!]
\centering
\includegraphics[scale=0.40]{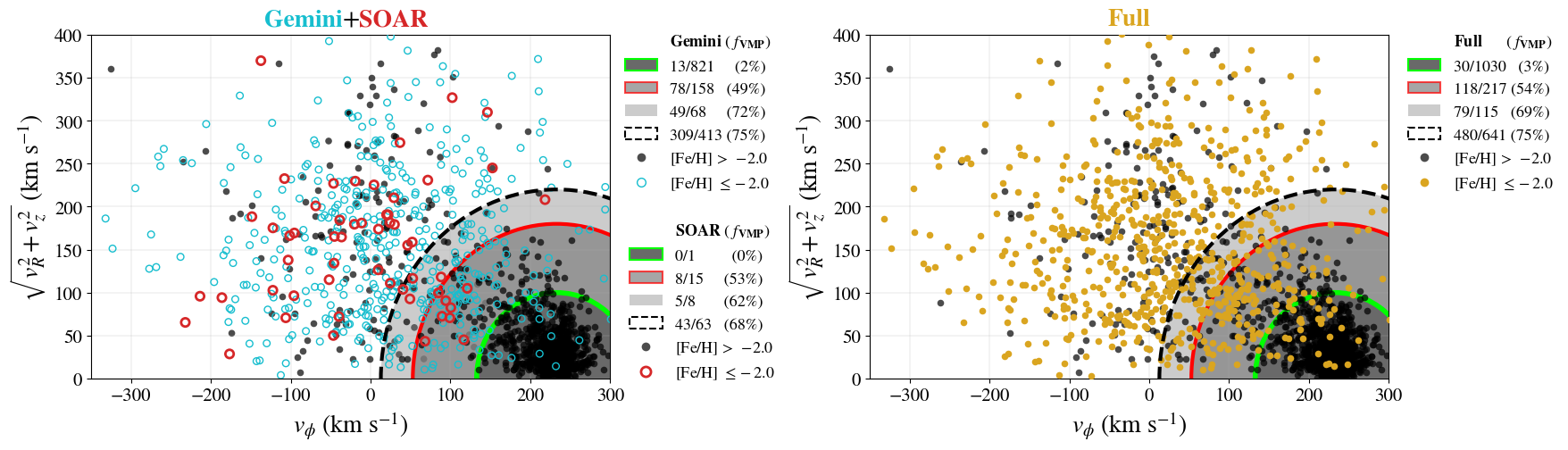}
\includegraphics[scale=0.40]{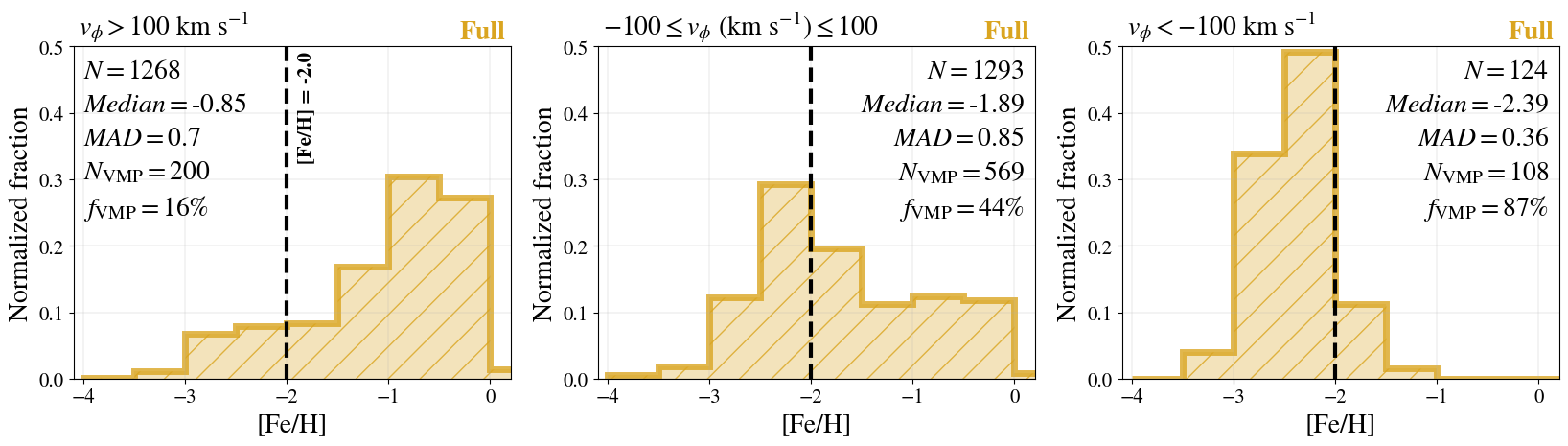}
\caption{{Top panels: Toomre diagrams ($\sqrt{v_R^2 + v_z^2}$ vs. $v_{\phi}$). Top left: Gemini (Section \ref{gemini}) and SOAR (Section \ref{soar}) samples. Top right: Full (Gemini+SOAR+\citetalias{placco2019}) sample. Stars that are VMP are represented with colored symbols in the same color scheme as Figures \ref{fig:projs}, \ref{fig:TeffLogg}, \ref{fig:abunds}, \ref{fig:hists}, and \ref{fig:santucci}. Black dots are stars with [Fe/H] $>-2.0$. The green and red solid lines mark the velocity boundaries for the thin ($|V_{\rm total}-V_{\rm LSR}| < 100$ km s$^{-1}$; dark gray) and thick disks (100~km~s$^{-1} \leq |V_{\rm total}-V_{\rm LSR}| < 180$~km~s$^{-1}$; medium gray), respectively (see text). The dashed line marks the transition region (180~km~s$^{-1} \leq |V_{\rm total}-V_{\rm LSR}| < 220$~km~s$^{-1}$; light gray stripe) between the thick disk and halo. Finally, the halo is defined by $|V_{\rm total}-V_{\rm LSR}| \geq 220$ km s$^{-1}$ (white region). To the right of each panel, the specific VMP fractions ($f_{\rm VMP}$) in each of the described areas in ($\sqrt{v_R^2 + v_z^2}$, $v_{\phi}$) space are provided. Bottom panels: Metallicity distribution functions (MDFs) of the full sample for various intervals of $v_{\phi}$. Bottom left: $v_{\phi} > 100$ km s$^{-1}$. Bottom middle: $-100 \leq v_{\phi} \rm \ (km \ s^{-1}) \leq 100$. Bottom right: $v_{\phi} < -100$ km s$^{-1}$. The total numbers of stars ($N$), the medians, median absolute deviations (MAD), the numbers ($N_{\rm VMP}$), and the fractions ($f_{\rm VMP}$) of VMP stars are also displayed in the corresponding panels. The dashed vertical lines mark the VMP limit at [Fe/H] = $-2.0$, for reference. }
\label{fig:toomre}} 
\end{figure*}

\subsection{Toomre Diagrams}
\label{toomre}

For comparison, we have also investigated the success rates in finding low-metallicity stars using a kinematic-based target selection within different regions of the Toomre diagram (Figure \ref{fig:toomre}). The peculiar motion of the Sun with respect to the local standard of rest (LSR) is $(U,V,W)_{\odot}$~$=~(11.10,12.24,7.25)$ km s$^{-1}$ \citep{schon2010}. The adopted velocity of the LSR is $V_{\rm LSR}~=~(0.0,232.8,0.0)$~km~s$^{-1}$ \citep{mcmillan2017}. Recently, purely acceleration-based estimates of the Galactic fundamental parameters have been made available, thanks to the direct measurement of the Solar System's acceleration from Gaia EDR3 astrometry. Nevertheless, the values adopted here have been chosen due to their consistency with the Galactic model of \citet{mcmillan2017}, which will be employed for orbit integration in Section \ref{sec:orbits}. Figure \ref{fig:toomre} shows the $(\sqrt{ v_R^2 + v_z^2}, v_{\phi})$ space, where ($v_R$, $v_{\phi}$, $v_z$) compose the velocity vector of each star in the cylindrical coordinate system (radial, azimuthal, and vertical directions, respectively). Stars in prograde motion have $v_{\phi} > 0$. Note that it is necessary to have measured $V_{\rm los}$ in order to calculate the complete set of ($v_R$, $v_{\phi}$, $v_z$). This is a clear disadvantage in utilizing the Toomre diagram for selecting targets for systematic searches of low-metallicity stars, since many targets would be ignored due to lack of available $V_{\rm los}$. Therefore, we wish to evaluate whether or not the efficiency in finding VMP stars is hindered by the application of $Z_{\rm Gal}$ vs. $V_{\rm TAV}$ diagram in comparison to $\sqrt{v_R^2 + v_z^2}$ vs. $v_{\phi}$, where the complete phase-space vector needs to be available. In this context, we also study changes in the full sample's MDF with $v_{\phi}$ (see Figure \ref{fig:toomre}).

For consistency with the analysis presented in Section \ref{sec:santucci}, we have divided the Toomre diagrams into four regions, similar to what has been done for the ($Z_{\rm Gal}$, $V_{\rm TAV}$) space. In Figure \ref{fig:toomre}, these are displayed in an arrangement similar to Figure \ref{fig:santucci}, with a gray scale tentatively representing different Galactic components. Stars that are VMP are shown with colored symbols in the same color scheme as Figures \ref{fig:projs}, \ref{fig:TeffLogg}, \ref{fig:abunds}, \ref{fig:hists}, and \ref{fig:santucci}. The dark gray area with a green contour contains stars within $|V_{\rm total}-V_{\rm LSR}| < 100$~km~s$^{-1}$, where $V_{\rm total}$ is the complete velocity vector of a given star. This is the most crowded zone in both top panels of Figure \ref{fig:toomre}, which is noticeable from the accumulation of black points around $v_{\phi} \sim V_{\rm LSR}$. This region in velocity space is representative of the thin disk (e.g., \citealt{Venn2004, Bensby2014}), and is also recognizable from its $f_{\rm VMP} \lesssim 3\%$ in both diagrams of Figure \ref{fig:toomre}, despite the B\&B selection. The predominance of this component is also perceptible from the MDF of stars with $v_{\phi}~>~100$~km~s$^{-1}$ (bottom left panel of Figure \ref{fig:toomre}). This MDF peaks at [Fe/H] $\sim -0.8$, but also shows an extended VMP tail associated with the thick disk and prograde portion of the halo.

Moving radially outwards from the thin-disk region of the $(\sqrt{ v_R^2 + v_z^2}, v_{\phi})$ space, the medium gray area in between the green and red lines is defined by 100~km~s$^{-1} \leq |V_{\rm total}-V_{\rm LSR}| < 180$~km~s$^{-1}$, and is characteristic of the thick disk (see, e.g., \citealt{amarante2020b} for a recent discussion). A noteworthy feature of the Toomre diagrams in Figure \ref{fig:toomre} is the presence of high fractions ($\gtrsim$50\%) of VMP stars in this thick-disk-like slice of the velocity space. We speculate that this feature is also due to the underlying B\&B selection. This finding has led us to scrutinize the presence of VMP and EMP stars associated with the Galactic disks further in Section \ref{sec:orbits}.

The light gray stripe, defined by the dashed and red solid lines in Figure \ref{fig:toomre} (180~km~s$^{-1} \leq |V_{\rm total}-V_{\rm LSR}| < 220$~km~s$^{-1}$), comprises the tail of the velocity distribution of the Galactic disk (e.g., \citealt{bonaca2017} and \citealt{koppelman2018}). This transition region between thick disk and halo also contains stars from the ``splashed disk" \citep{belokurov2020, Amarante2020a, An2020}, considered to be the metal-rich, \textit{in-situ} counterpart of the halo \citep{bonaca2017,Bonaca2020,Haywood2018, gallart2019, DiMatteo2019}. 
The $f_{\rm VMP}$ in this area of the $(\sqrt{ v_R^2 + v_z^2}, v_{\phi})$ space is 
statistically equivalent to the halo selection.  

We note that the concentration of stars within $-50 \lesssim v_{\phi}$ (km s$^{-1}$) $\lesssim +50$, permeating all values of $\sqrt{v_R^2 + v_z^2}$, is reminiscent of the Gaia-Sausage/Enceladus (GSE) merging event \citep{belokurov2018, Haywood2018, helmi2018}. However, considering $v_{\phi}$ alone, we also note the extended MDF shown in the bottom middle panel of Figure \ref{fig:toomre} ($-100 \leq v_{\phi} \rm \ (km \ s^{-1}) \leq 100$), which peaks between $-2.5 < \rm [Fe/H] < -2.0$, but has a prominent metal-rich tail linked to the thick and splashed disks. On the other hand, the MDF in the $v_{\phi} < -100$ km s$^{-1}$ panel of Figure \ref{fig:toomre} (bottom right) is almost entirely comprised by VMP stars. These stars might also be associated with previously identified halo substructures, e.g., Sequoia \citep{myeongSequoia} and Thamnos \citep{koppelman2019}. Indeed, independent efforts have suggested that these retrograde groups are more metal-poor than GSE \citep{matsuno2019, Monty2020, dietz2020, Kordopatis2020}. Nevertheless, these supposed remnants of now-destroyed dwarf galaxies are also of interest to this work, particularly because, recently, large numbers of VMP stars, including those exhibiting $r$-process-element enhancement \citep{Gudin2021}, have been demonstrated to be associated with them (\citealt{yuan2020, Limberg2021}; see Section \ref{sec:orbits}). 

Finally, the black dashed line delineates the $|V_{\rm total}-V_{\rm LSR}| \geq 220$ km s$^{-1}$ zone of the Toomre diagrams. These are the white areas in both top panels of Figure \ref{fig:toomre}. Stars inhabiting this part of the $(\sqrt{v_R^2 + v_z^2}, v_{\phi})$ space are from either the \textit{in-situ} or accreted components of the halo. The $f_{\rm VMP}$ of the full sample in this white region is similar to the one found for the halo-like portions of the $Z_{\rm Gal}$ vs. $V_{\rm TAV}$ diagram in Section \ref{sec:santucci}. Nonetheless, we have found no statistical differences in $f_{\rm VMP}$ between the prograde and retrograde components of the halo. To conclude, kinematic criteria based on the ($Z_{\rm Gal}$, $V_{\rm TAV}$) space can be as efficient as the traditional Toomre diagram in finding low-metallicity stars, with the additional advantage of not needing prior measurement of $V_{\rm los}$, thus enabling access to a larger number of candidates, while making optimal usage of this parameter whenever it is available. Thus, we recommend consideration of the $V_{\rm TAV}$ quantity for the target selection of observational campaigns aiming to find VMP and EMP stars in the Galaxy.

\subsection{Orbits of VMP/EMP Stars}
\label{sec:orbits}

We now examine the presence of dynamically interesting stars in all samples, which will become priority targets for high-resolution follow-up. First, we check for VMP and EMP stars residing in the kinematically defined thick disk (Section \ref{toomre}). In order to confirm whether or not these stars are truly confined to the disk system, we have integrated their orbits with the publicly available library \texttt{AGAMA} \citep{agama} forward in time for $\sim$5 Gyr. The Galactic model employed is from \citet{mcmillan2017}, and includes stellar and gaseous disks, a flattened bulge, and a spherical dark matter halo. The assumptions regarding the Galactic fundamental parameters are the same as Sections \ref{sec:kinematics} and \ref{toomre}. The $V_{\rm los}$ have been acquired from Gaia EDR3, where available, but we have only considered stars with \texttt{parallax\_over\_error} $\geq 5$ for this exercise. We have performed 1000 realizations of each star's orbit according to its uncertainties in the astrometric quantities, assuming Gaussian distributions for them. The medians of each dynamical parameter have been taken as our nominal values. Our investigation is focused on the maximum vertical distance achieved during a star's orbit ($Z_{\max}$), the perigalactic distance ($r_{\min}$), the apogalactic distance ($r_{\max}$), eccentricity ($e = (r_{\max}-r_{\min})/(r_{\max}+r_{\min})$), orbital energy ($E$), and the vertical component of angular momentum $L_z = R_{\rm Gal} \times v_{\phi}$, where $R_{\rm Gal} = \sqrt{X_{\rm Gal}^2 + Y _{\rm Gal}^2}$ is the plane-projected distance of a given star from the Galactic center. 

\begin{figure}[!ht]
\centering
\includegraphics[width=\columnwidth]{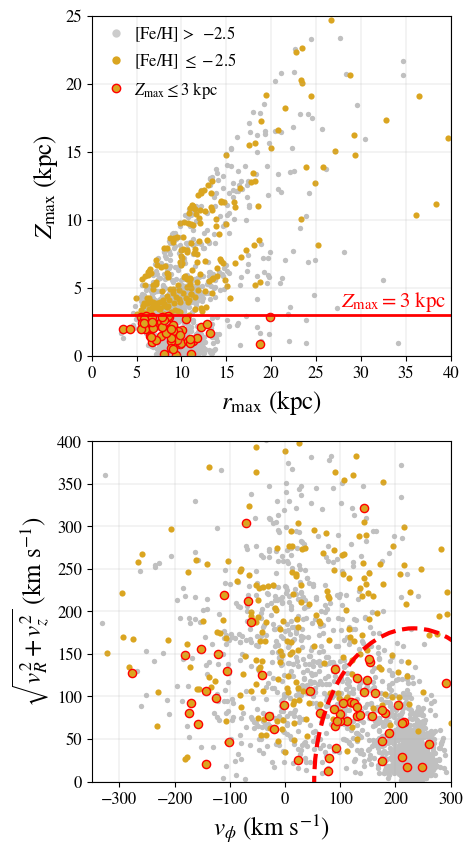}
\caption{{Top panel: $Z_{\max}$ vs. $r_{\max}$. The red horizontal line at $Z_{\max} = 3$ kpc marks the limit for stars considered to be on planar or halo-like orbits in our analysis (see text). Bottom panel: Toomre diagram, similar to Figure \ref{fig:toomre}. The red dashed line marks the $|V_{\rm total}-V_{\rm LSR}| = 180$ km s$^{-1}$ boundary between the thick disk and the transition region as in Section \ref{toomre}. In both panels, gray dots represent stars with metallicities [Fe/H] $> -2.5$. Stars with [Fe/H] $\leq-2.5$ are displayed as yellow dots. Among these, the planar ones are plotted with red edges.}
\label{fig:ZmaxRmax}} 
\end{figure}

The top panel of Figure \ref{fig:ZmaxRmax} shows the distribution of the full sample in the $Z_{\max}$ vs. $r_{\max}$ space. A star is considered to be on a planar orbit if its $Z_{\max} \leq 3$ kpc. The transition between the thick disk and halo is smooth in this diagram, as already noticed from the full sample's MDFs for different slices of both $|Z|_{\rm Gal}$ (Figure \ref{fig:hists}) and $v_{\phi}$ (Figure \ref{fig:toomre}). However, the $Z_{\max}$ limit at 3 kpc reflects the recently estimated scale height of the thick disk by \citet{LiZhao2017}. Crucially, it is also consistent with what has been considered in the analyses of \citealt{sestito2019, sestito2020}. Interestingly, a separation between thick disk and halo at $Z_{\max} \sim 3$ kpc has also been achieved by \citet[]{PerezVillegas2020} in their independent analysis of orbits of Galactic globular clusters. We have paid particular attention to stars with [Fe/H] $\leq -2.5$, as these are potential members of the aforementioned mono-enriched class \citep{hartwig2018}. Hence, this metallicity cutoff is, apparently, more physically significant than the classic definitions of VMP and EMP stars, and is also consistent with the analyses of \citet{sestito2020, Sestito2021}. For convenience, throughout this section, we refer to a star as ``VMP/EMP" when [Fe/H] $\leq -2.5$. A total of 275 stars in the full sample are consistent with this metallicity regime. These are shown as yellow dots in Figure \ref{fig:ZmaxRmax}, while those with [Fe/H] $>-2.5$ are represented in gray. All VMP/EMP stars on planar orbits are plotted with red edges, and correspond to a fraction of $22^{+5}_{-4}\%$ (60/275), consistent with both \citet{sestito2020} and \citet{Cordoni2020}, despite our small biases towards halo kinematics (Section~\ref{sec:obs}).

The bottom panel of Figure \ref{fig:ZmaxRmax} presents the Toomre diagram for the full sample, divided according to the prescription above. One can see that there is a substantial number of VMP/EMP stars on low-$Z_{\max}$ orbits occupying the region of the $(\sqrt{ v_R^2 + v_z^2}, v_{\phi})$ space characteristic of the thin and thick disks, as conjectured in Sections \ref{sec:santucci} and \ref{toomre}. These stars can be considered part of the third stellar population that resides in the Galactic disk, known as the metal-weak thick disk (MWTD; \citealt{Norris1985, morrison1990, Beers1995, Bonifacio1999, chiba2000, Beers2002, beers2014, Brown2008, Reddy2008, Kordopatis2013, carollo2014,carollo2019,  Hawkins2015,  LiZhao2017,LiZhao2018, An2020, DiMatteo2020}). Despite early suggestions that the MDF of the MWTD might extend down to the VMP/EMP regime \citep{Beers1995, Bonifacio1999}, confirmation with large numbers of VMP/EMP stars was only possible thanks to the powerful combination of Gaia DR2 with large-scale spectroscopic surveys (e.g., LAMOST; \citealt{sestito2020}). In fact, there is increasing evidence for the existence of an ``ultra"-MWTD, reaching metallicities $<-4.0$ \citep{sestito2019, Cordoni2020, DiMatteo2020}. An UMP star has even been found in the thin disk ($Z_{\max} < 0.5$ kpc; $e < 0.2$)  \citep{Schlaufman2018}. On the other hand, about a third of the planar VMP/EMP stars are rotating counter to the Galactic disks, some of them with $v_{\phi} \lesssim -100$ km s$^{-1}$).  

\begin{figure}[!ht]
\centering
\includegraphics[width=\columnwidth]{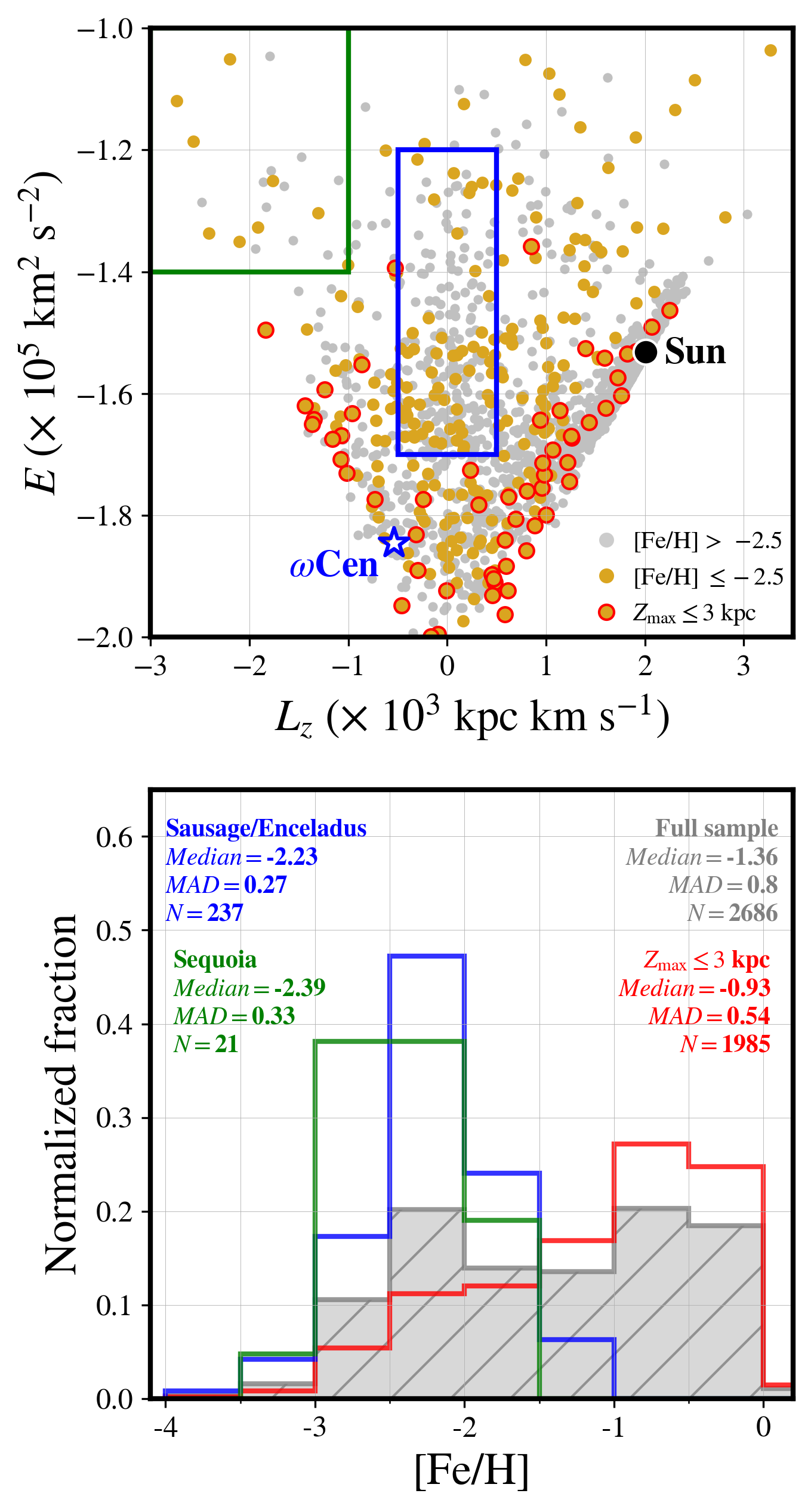}
\caption{{Top panel: $E$ vs. $L_z$. The position of the Sun is indicated with a black circle: $(E,L_z)_{\odot} = (-1.5 \times 10^5 \ \rm km^2 \ s^{-2}, +2.0 \times 10^3 \ kpc \ km \ s^{-1})$. The blue solid lines delineate our selection box for Gaia-Sausage/Enceladus (GSE; see text) stars. The blue star shows the position of the $\omega$ Centauri globular cluster: $(E,L_z)_{\omega \rm Cen} = (-1.8 \times 10^5 \ \rm km^2 \ s^{-2}, -0.5 \times 10^3 \ kpc \ km \ s^{-1})$. The green box represents the selection of Sequoia stars (see text). Gray dots represent stars with [Fe/H] $> -2.5$. Stars with [Fe/H] $\leq-2.5$ are displayed as yellow dots. Among these, the disk-like stars are plotted with red edges. Bottom panel: Metallicity distribution functions of the full (gray), the planar (red), GSE (blue), and Sequoia (green) samples. The medians, the median absolute deviations (MAD), and the total numbers of stars ($N$) are also shown.}
\label{fig:ExLz}} 
\end{figure}

The VMP/EMP stars on planar orbits are of crucial importance to constrain the early formation of the Galaxy. Recent cosmological zoom-in simulations have been pointing to the majority of VMP/EMP stars being born very early on ($z \gtrsim 5$), but mostly \textit{ex-situ}, in primordial galaxies that have been accreted by the main progenitor through hierarchical assembly up until $z \sim 1$ \citep{Starkenburg2017SIMULATIONS,El-Badry2018}. Hence, these objects would be deposited into the main halo with a variety of different orbits, either prograde or retrograde \citep{Sestito2021}, and become phase-space mixed with those of \textit{in-situ} origin. However, given the presence of the ultra-MWTD, indistinguishable from the canonical thick disk from kinematics alone, \citet{DiMatteo2020} emphasize that the simplest deduction would be that these VMP/EMP stars were born in the Galaxy itself. Hence, these stars would have suffered from the same heating mechanism that operated to form both the splashed disk and the thick disk, i.e., the GSE event \citep{DiMatteo2019, gallart2019, belokurov2020,Bonaca2020}, although the exact role that merging events play in dynamical heating remains under debate (see, e.g., \citealt{Jean-Baptiste2017} and \citealt{Amarante2020a}). Nevertheless, these scenarios are not necessarily mutually exclusive, although the retrograde fraction of the VMP/EMP stars on planar orbits might be difficult to reconcile with the latter. In fact, \citet{Sestito2021} has noted that the Galaxy apparently hosts an excess of retrograde, low-$Z_{\max}$, low-metallicity stars in comparison to their simulations. The authors have further suggested that this aspect might be intrinsically related to a Milky Way-specific evolution, particularly the  accretion of GSE's progenitor system. Such features of VMP/EMP stars on planar orbits in the Galaxy, especially the prograde/retrograde ratio, might help us understand this puzzling assembly history.

We continue our exploration for dynamically interesting stars in Figure \ref{fig:ExLz}, where the $E$ vs. $L_z$ diagram is presented (top panel). This parameter space has been extensively utilized in the search for substructures in the Galaxy, especially in the halo (e.g., \citealt{chiba2000} and numerous subsequent authors; see \citealt{Helmi2020} for a recent review). Unsurprisingly, we notice the spread of these objects on planar orbits throughout the locus, in $(E,L_z)$ space, associated with either the disks or the low-$E$ ($\lesssim -1.6 \times 10^5$ km$^2$ s$^{-2}$), retrograde halo. Curiously, the latter is also commonly attributed to the remnant of the proposed Thamnos event \citep{koppelman2019}. Indeed, we had already noticed that VMP stars from this substructure are unequivocally contained within $Z_{\max} \lesssim 3.5$ kpc in \citet{Limberg2021}. One might even speculate that the accretion of a small system like the Thamnos progenitor could have been the responsible for the apparent excess of retrograde low-$Z_{\max}$ stars reported by \citet[]{Sestito2021}. However, this should be taken with caution, since both the specific chemical profile of this substructure \citep{Monty2020}, as well as its independence from GSE \citep{Kordopatis2020}, are still under active study (see also discussions about the consequences of massive mergers in \citealt{Jean-Baptiste2017} and \citealt{Koppelman2020MassiveMerger}).

From the $(E,L_z)$ space, we have also tentatively isolated stars from both GSE and Sequoia debris. We have placed characteristic selection boxes for each of these substructures (blue and green lines, respectively, in the top panel of Figure \ref{fig:ExLz}), in an approach similar to both \citet[who analyzed the MDFs and $\alpha$-to-iron ratios of their member stars]{matsuno2019} and \citet[who studied possible dynamical associations between globular clusters and various Galactic components]{massari2019}. For reference, we have also located the $\omega$ Centauri globular cluster in $(E,L_z)$ space, which has been proposed to be the remnant nuclear star cluster of GSE \citep{Pfeffer2021}. The GSE selection is: $-1.7 \leq E \ (\times 10^5 \ \rm km^2 \ s^{-2}) \leq -1.2$, and $-0.5 \leq L_z \ (\times 10^3 \ \rm kpc \ km \ s^{-1}) \leq +0.5$. For Sequoia: $-1.4 \leq E \ (\times 10^5 \ \rm km^2 \ s^{-2}) \leq -1.0$, and $L_z \leq -1.0 \times 10^3$ kpc km s$^{-1}$. The MDFs resulting from these cuts are presented in the bottom panel of Figure \ref{fig:ExLz} with the same colors as their respective selection boxes. For comparison, we have also plotted the MDFs for the full sample (gray) and for those stars on planar orbits (red). The MDFs from both GSE and Sequoia mostly occupy the metallicity range of $-3.0 < [\rm Fe/H] < -1.5$, with medians of [Fe/H] $\sim -2.2$ and $\sim -2.4$, respectively, but median absolute deviations that make them compatible with each other. Despite some overlap, these are much more metal poor than the typical values found for these substructures in the literature (e.g., \citealt{helmi2018, Conroy2019B, Vincenzo2019, amarante2020b, Feuillet2020} for GSE and \citealt{myeongSequoia, matsuno2019,Monty2020,dietz2020} for both). We have experimented with other selections in velocity space, as recommended by \citet[]{Feuillet2020}, but the results are equivalent. We conclude that this low-metallicity bias is an effect of the underlying B\&B selection.

From the bottom panel of Figure \ref{fig:ExLz}, we note that the MDF of GSE is reminiscent of those presented in Figure \ref{fig:hists} (for $|Z|_{\rm Gal} > 3$ kpc and $V_{\rm TAV} > 150$ km s$^{-1}$), commensurate with this substructure being the predominant population in the nearby, accreted halo. The presence of this component is also noticeable from the behavior of the full sample's MDF (Figure \ref{fig:ExLz}), which has two peaks, with the lowest-metallicity one being in the same location as GSE's ($-2.5 \leq [\rm Fe/H] \leq -2.0$). The other peak, at $-1.0 \leq [\rm Fe/H] \leq -0.5$, is associated with metal-rich, thin-disk stars, mistakenly selected as VMP candidates and followed-up during our program. We note, however, the extended low-metallicity tail of the MDF representing the low-$Z_{\max}$ portion of the full sample, related to the previously discussed VMP/EMP stars with thick-disk-like kinematics. 

Finally, we stress that not only the VMP/EMP stars with $Z_{\max} > 3$ kpc, but also those confined to the Galactic plane, exhibit a wide variety of orbital behaviors, as seen in Figures \ref{fig:ZmaxRmax} and \ref{fig:ExLz}. Both of these populations span the full range of orbital $e$, a feature highlighted by \citet[]{Cordoni2020} in their analysis, from very low ($e < 0.2$), disk-like, to very high ($e > 0.8$), typical of GSE debris. The mild-$e$ ($0.4 < e < 0.6$) stars might also be associated with substructures of \textit{ex-situ} nature (see \citealt{Helmi2020}), but most of them apparently belong to the \textit{in-situ} counterpart of the halo. \citet[]{Cordoni2020} highlights that VMP/EMP stars presenting low-to-mild-$e$, low-$Z_{\max}$ orbits follow the $e$ distributions of simulated thick disks constructed via heating mechanisms by \citet[]{Sales2009}. The retrograde counterpart of such a population would likely be an accreted one according to these authors. This presented ``continuity" (in the words of \citealt{DiMatteo2020}) in the dynamical properties of the most metal-poor stars will certainly need to be taken into consideration in future simulation efforts trying to reproduce the birth and evolution of the Milky Way.

\section{Conclusions}
\label{sec:conc}

In this paper, we have presented a low-resolution ($R \sim 2000$) spectroscopic study of 1897 metal-poor star candidates selected in the Best \& Brightest Survey \citep{beb2014}. The observations have been conducted with either the GMOS/Gemini (North and South) or Goodman/SOAR combinations between semesters 2014A and 2019B. We have obtained the atmospheric parameters ($T_{\rm eff}$, $\log g$, and [Fe/H]) for these stars, as well as carbon and magnesium abundance ratios for most of them. Furthermore, we have utilized the phase-space information provided by Gaia EDR3 to perform an in-depth investigation on the influence of kinematic-based target selection criteria in the efficiency of finding VMP stars. Finally, we have explored the presence of dynamically interesting VMP/EMP stars in our sample. The main results can be summarized as follows.

$\bullet$ Overall, $56\%$ (1064) of the newly observed stars have [Fe/H] $< -1.0$, $30\%$ (566) are VMP, and $2\%$ (35) are EMP. Combined with the previously published data of \citetalias{placco2019}, the full B\&B sample is now one the largest homogeneously-analyzed compilations of bright ($V \lesssim 14$) VMP and EMP stars available.   

$\bullet$ There are 191 CEMP stars in the Gemini+SOAR sample (94 from Group I and 97 from Group II in the Yoon-Beers $A$(C)$-$[Fe/H] diagram).

$\bullet$ The fraction of CEMP stars increases with decreasing metallicity. In the VMP and EMP regimes, we have found $19\%$ and $43\%$, respectively.

$\bullet$ We have introduced the $V_{\rm TAV}$ quantity, which demands either PMs or $V_{\rm los}$, but makes optimal use of their combination when both are available. We have explored the $Z_{\rm Gal}$ vs. $V_{\rm TAV}$ diagrams. and confirmed that the $f_{\rm VMP}$ grows for greater distances from the Galactic plane and velocities within the full B\&B sample. 

$\bullet$ The final success rates that have been achieved for $Z_{\rm Gal} \geq 0.5$ kpc and $V_{\rm TAV} \geq 100$ km s$^{-1}$ are $96\%$, $76\%$, $28\%$, and $4\%$ for [Fe/H] $\leq-1.5$, $\leq-2.0$, $\leq-2.5$, and $\leq-3.0$, respectively. 

$\bullet$ Using the Toomre diagram, the $f_{\rm VMP}$ in the kinematically-defined halo is equivalent (at the $1\sigma$ level) to what has been found for the most interesting regions of the ($Z_{\rm Gal},V_{\rm TAV}$) space. 

$\bullet$ After integrating their orbits, $22\%$ of the stars with [Fe/H] $\leq -2.5$ have been found to be confined within $Z_{\max} \leq 3$ kpc. However, the VMP/EMP stars vetted here exhibit a wide variety of orbital behaviors, spanning all values of $L_z$ and orbital eccentricity, in keeping with the post-Gaia literature.

$\bullet$ Most of the VMP/EMP stars on planar orbits can be kinematically (100~km~s$^{-1} \leq |V_{\rm total}-V_{\rm LSR}| < 180$~km~s$^{-1}$) and dynamically ($e \lesssim 0.6$) attributed to the Galactic metal-weak thick disk. Moreover, a third of these planar, mild-$e$ stars are retrograde (some with $v_{\phi} \lesssim -100$ km s$^{-1}$), and their origin remains unclear. 

$\bullet$ Stars on halo-like orbits ($Z_{\max} > 3$ kpc) with $e \gtrsim 0.8$ are generally associated with the GSE merging event. On the other hand, those with mild values of $e$ might be linked to either the \textit{in-situ} counterpart of the halo or other accreted substructures (e.g., Sequoia). 

We note that there remain many thousands of candidate metal-poor stars originally identified in the B\&B survey, as well as in the HK and Hamburg/ESO surveys, which have not yet been vetted with low-resolution spectroscopy.  Given the high success rates for the identification of VMP stars demonstrated by our kinematic-selection approach, it is feasible to go directly to high-resolution follow-up for many of these stars in future campaigns.

\acknowledgments
The authors thank Angeles Pérez-Villegas for insightful discussions at the early stages of this work. We also thank the anonymous referee for a useful report. G.L. acknowledges CAPES (PROEX; Proc. 88887.481172/2020-00) and CNPq (PIBIC; Proc. 144638/2018-5). R.M.S. acknowledges CNPq (Proc. 436696/2018-5 and 306667/2020-7). S.R. would like to acknowledge support from FAPESP (Proc. 2015/50374-0 and 2014/18100-4), CAPES, and CNPq. H.D.P. thanks FAPESP Proc. 2018/21250-9. D.S., V.M.P., and T.C.B.
acknowledge partial support from grant PHY 14-30152, Physics Frontier Center/JINA Center for the Evolution of the Elements (JINA-CEE), awarded by the US National Science Foundation. The work of V.M.P. is supported by NOIRLab, which is managed by the Association of Universities for Research in Astronomy (AURA) under a cooperative agreement with the National Science Foundation. Y.S.L. acknowledges support from the National Research Foundation (NRF) of Korea grant funded by the Ministry of Science and ICT (NRF-2018R1A2B6003961).  

This work is partially based on observations obtained under the programs SO-2018B-010 and SO-2019B-013 at the Southern Astrophysical Research (SOAR) telescope, which is a joint project of the Minist\'{e}rio da Ci\^{e}ncia, Tecnologia e Inova\c{c}\~{o}es (MCTI/LNA) do Brasil, the US National Science Foundation’s NOIRLab, the University of North Carolina at Chapel Hill (UNC), and Michigan State University (MSU). This research is also partially based on observations obtained under programs GN-2015A-Q-76, GN-2015B-Q-86, GN-2016A-DD-3, GN-2016A-Q-75, GN-2016B-Q-77, GN-2017A-Q-82, GN-2017B-Q-75, GN-2017B-Q-79, GN-2018A-Q-403, GN-2018B-Q-316, GN-2019A-Q-309, GN-2019B-Q-403, GS-2014A-Q-74, GS-2014A-Q-8, GS-2015A-Q-77, GS-2015A-Q-92, GS-2015B-Q-71, GS-2016A-Q-76, GS-2016B-Q-81, GS-2017A-Q-86, GS-2017B-Q-75, GS-2017B-Q-84, GS-2018A-Q-406, and GS-2018B-Q-315 at the international Gemini Observatory, a program of NSF’s NOIRLab, which is managed by the Association of Universities for Research in Astronomy (AURA) under a cooperative agreement with the National Science Foundation on behalf of the Gemini Observatory partnership: the National Science Foundation (United States), National Research Council (Canada), Agencia Nacional de Investigaci\'on y Desarrollo (Chile), Ministerio de Ciencia, Tecnolog\'ia e Innovaci\'on (Argentina), Minist\'{e}rio da Ci\^{e}ncia, Tecnologia e Inova\c{c}\~{o}es (Brazil), and Korea Astronomy and Space Science Institute (Republic of Korea).

This work has made use of data from the European Space Agency (ESA) mission {\it Gaia} (\url{https://www.cosmos.esa.int/gaia}), processed by the {\it Gaia} Data Processing and Analysis Consortium (DPAC, \url{https://www.cosmos.esa.int/web/gaia/dpac/consortium}. Funding for the DPAC has been provided by national institutions, in particular the institutions participating in the {\it Gaia} Multilateral Agreement. This research has made use of the SIMBAD database and VizieR catalogue access tool, operated at CDS, Strasbourg, France. This research was made possible through the use of the AAVSO Photometric All-Sky Survey (APASS), funded by the Robert Martin Ayers Sciences Fund and NSF AST-1412587. This publication makes use of data products from the Two Micron All Sky Survey, which is a joint project of the University of Massachusetts and the Infrared Processing and Analysis Center/California Institute of Technology, funded by the National Aeronautics and Space Administration and the National Science Foundation. 

This research has been conducted despite the ongoing dismantling of the Brazilian scientific system. 

\smallskip

\facilities{Gemini North (8.1 m): GMOS-N, Gemini South (8.1 m): GMOS-S, SOAR (4.1 m): Goodman}

\software{
          {\tt matplotlib} \citep{matplotlib},
          {\tt Numpy} \citep{numpy},
          {\tt scipy} \citep{scipy}.
          }

\bibliographystyle{aasjournal}

\newpage

\bibliography{bibliography.bib}

\clearpage

\setcounter{table}{0}


\renewcommand{\arraystretch}{1.0}
\setlength{\tabcolsep}{0.47em}

\begin{longtable}{>{\footnotesize}c >{\footnotesize}c >{\footnotesize}c >{\footnotesize}r >{\footnotesize}c >{\footnotesize}r >{\footnotesize}c >{\footnotesize}c >{\footnotesize}c >{\footnotesize}c >{\footnotesize}c}
\caption{Coordinates and Observing Details} \\
\toprule
Star Name & Star Name & R.A. & \multicolumn{1}{c}{Decl.} & $l$ & \multicolumn{1}{c}{$b$} & Telescope & Instrument & Program ID    \\
(2MASS) & (Gaia EDR3) & (deg) & \multicolumn{1}{c}{(deg)} & (deg)    & \multicolumn{1}{c}{(deg)} & & & \\ \midrule \midrule
\endfirsthead
\multicolumn{9}{c}%
{{\bfseries Table \thetable\ }\textit{(continued)}} \\
\toprule
Star Name & Star Name & R.A. & \multicolumn{1}{c}{Decl.} & $l$ & \multicolumn{1}{c}{$b$} & Telescope & Instrument & Program ID    \\
(2MASS) & (Gaia EDR3) & (deg) & \multicolumn{1}{c}{(deg)} & (deg)    & \multicolumn{1}{c}{(deg)} & & & \\ \midrule \midrule
\endhead
\hline
\multicolumn{9}{c}{\textit{(continued)}}\\
\endfoot
\endlastfoot

00003305-7953389 & 4634573766005607552      & 0.1375       & -79.8942      & 305.7037 & -36.9587 & SOAR         & Goodman    & SO-2019B-013    \\
00020162-4430117 & 4994519032163925632      & 0.5067       & -44.5033      & 329.4221 & -69.9972 & SOAR         & Goodman    & SO-2019B-013    \\
00040305-6106367 & 4905632480654004608      & 1.0127       & -61.1102      & 312.9121 & -55.0903 & Gemini South & GMOS-S     & GS-2016A-Q-76   \\
00043646+4124062 & 384060304935385984       & 1.1519       & 41.4017       & 113.5728 & -20.6174 & Gemini North & GMOS-N     & GN-2017B-Q-75   \\
00045403+3524010 & 2876804519751163008      & 1.2251       & 35.4003       & 112.3464 & -26.5131 & Gemini North & GMOS-N     & GN-2016A-Q-75   \\
00052731+0025510 & 2546113345293028608      & 1.3638       & 0.4309        & 99.2097  & -60.3059 & Gemini North & GMOS-N     & GN-2015B-Q-86   \\
00060902-6238019 & 4904480471642380928      & 1.5376       & -62.6339      & 311.7034 & -53.7163 & Gemini South & GMOS-S     & GS-2015B-Q-71   \\
00071189+4724466 & 393031258963415936       & 1.8005       & 47.4133       & 115.2160 & -14.7946 & Gemini North & GMOS-N     & GN-2015B-Q-86   \\
\bottomrule
\end{longtable}

\clearpage

\renewcommand{\arraystretch}{1.0}
\setlength{\tabcolsep}{0.62em}

\begin{longtable}{>{\small}c >{\small}c >{\small}c >{\small}c >{\small}c >{\small}c >{\small}c >{\small}c >{\small}c }
\caption{Colors, Magnitudes and Reddening Estimates} \\
\toprule
Star Name             & Star Name                & $V$     & ($B-V$)      & $G$        & ($G_{\rm BP}-G_{\rm RP}$)        & $J$    & ($J-K$) & $E(B-V)$    \\
(2MASS) & (Gaia EDR3) & & & & & & &               \\ \midrule \midrule
\endfirsthead
\multicolumn{9}{c}%
{{\bfseries Table \thetable\ }\textit{(continued)}} \\
\toprule
Star Name             & Star Name                & $V$     & ($B-V$)      & $G$        & ($G_{\rm BP}-G_{\rm RP}$)        & $J$    & ($J-K$) & $E(B-V)$    \\
(2MASS) & (Gaia EDR3) & & & & & & &               \\ \midrule \midrule
\endhead
\hline
\multicolumn{9}{c}{\textit{(continued)}}\\
\endfoot
\endlastfoot

00003305-7953389 & 4634573766005607552      & 12.411 & 0.775  & 12.184   & 1.069    & 10.804 & 0.535 & 0.074    \\
00020162-4430117 & 4994519032163925632      & 12.631 & 0.807  & 12.397   & 1.067    & 11.007 & 0.567 & 0.011    \\
00040305-6106367 & 4905632480654004608      & 12.818 & 1.075  & 12.417   & 1.466    & 10.546 & 0.745 & 0.010    \\
00043646+4124062 & 384060304935385984       & 12.612 & 1.025  & 12.371   & 1.082    & 10.896 & 0.643 & 0.073    \\
00045403+3524010 & 2876804519751163008      & 12.174 & 0.698  & 11.925   & 1.011    & 10.576 & 0.513 & 0.063    \\
00052731+0025510 & 2546113345293028608      & 12.376 & 0.864  & 12.123   & 1.110    & 10.702 & 0.563 & 0.059    \\
00060902-6238019 & 4904480471642380928      & 11.897 & 0.971  & 11.618   & 1.232    & 10.034 & 0.638 & 0.012    \\
00071189+4724466 & 393031258963415936       & 10.278 & 0.743  & 11.285   & 0.910    & 9.206  & 0.348 & 0.118    \\
\bottomrule
\end{longtable}

\clearpage

\renewcommand{\arraystretch}{1.0}
\setlength{\tabcolsep}{0.80em}

\begin{longtable}{>{\normalsize}c >{\normalsize}c >{\normalsize}c >{\normalsize}r >{\normalsize}r >{\normalsize}r >{\normalsize}r >{\normalsize}c >{\normalsize}r }
\caption{Stellar Atmospheric Parameters and Abundances} \\
\toprule
Star Name & Star Name & $T_{\rm eff}$ & $\log g$  & [Fe/H] & [C/Fe] & [C/Fe]$_c$ & $A$(C)$_c$ & [Mg/Fe]  \\
(2MASS) & (Gaia EDR3) & (K) & (cgs) & & & & & \\ \midrule \midrule
\endfirsthead
\multicolumn{9}{c}%
{{\bfseries Table \thetable\ }\textit{(continued)}} \\
\toprule
Star Name & Star Name & $T_{\rm eff}$ & $\log g$  & [Fe/H] & [C/Fe] & [C/Fe]$_c$ & $A$(C)$_c$ & [Mg/Fe]  \\
(2MASS) & (Gaia EDR3) & (K) & (cgs) & & & & & \\ \midrule \midrule
\endhead
\hline
\multicolumn{9}{c}{\textit{(continued)}}\\
\endfoot
\endlastfoot

00003305-7953389 & 4634573766005607552 & 5280  & 2.06  & -1.93 & 0.13  & 0.17  & 6.67  & 0.38   \\
00020162-4430117 & 4994519032163925632 & 5036  & 2.00  & -1.81 & 0.30  & 0.37  & 6.99  & $\dots$  \\
00040305-6106367 & 4905632480654004608 & 4648  & 4.13  & -1.27 & 0.15  & 0.15  & 7.31  & -0.19  \\
00043646+4124062 & 384060304935385984  & 6068  & 3.51  & -1.29 & 0.95  & 0.95  & 8.09  & 0.14   \\
00045403+3524010 & 2876804519751163008 & 5455  & 3.52  & -2.59 & 0.50  & 0.50  & 6.34  & 0.06   \\
00052731+0025510 & 2546113345293028608 & 5397  & 4.41  & -0.25 & 0.36  & 0.36  & 8.54  & 0.37   \\
00060902-6238019 & 4904480471642380928 & 4810  & 1.41  & -2.33 & 0.26  & 0.73  & 6.82  & 0.01   \\
00071189+4724466 & 393031258963415936  & 5868  & 3.98  & -0.53 & 0.46  & 0.46  & 8.36  & 0.30   \\
\bottomrule
\end{longtable}

\clearpage

\renewcommand{\arraystretch}{1.0}
\setlength{\tabcolsep}{0.66em}

\begin{longtable}{>{\small}c >{\small}c >{\small}r >{\small}c >{\small}c >{\small}r >{\small}r >{\small}r >{\small}r}
\caption{Phase-space Information from Gaia EDR3} \\
\toprule
Star Name & Star Name & \multicolumn{1}{c}{$V_{\rm los}$} & $d_{\rm helio}$  & $\sigma_d$ & \multicolumn{1}{c}{PM$_{\rm R.A.}$} & \multicolumn{1}{c}{PM$_{\rm Decl.}$} & \multicolumn{1}{c}{$Z_{\rm Gal}$} & \multicolumn{1}{c}{$V_{\rm TAV}$}    \\
(2MASS) & (Gaia EDR3) & \multicolumn{1}{c}{(km s$^{-1}$)} & (kpc) & (kpc) & \multicolumn{1}{c}{(mas yr$^{-1}$)} & \multicolumn{1}{c}{(mas yr$^{-1}$)} & \multicolumn{1}{c}{(kpc)} & \multicolumn{1}{c}{(km s$^{-1}$)} \\ \midrule \midrule
\endfirsthead
\multicolumn{9}{c}%
{{\bfseries Table \thetable\ }\textit{(continued)}} \\
\toprule
Star Name & Star Name & \multicolumn{1}{c}{$V_{\rm los}$} & $d_{\rm helio}$  & $\sigma_d$ & \multicolumn{1}{c}{PM$_{\rm R.A.}$} & \multicolumn{1}{c}{PM$_{\rm Decl.}$} & \multicolumn{1}{c}{$Z_{\rm Gal}$} & \multicolumn{1}{c}{$V_{\rm TAV}$}    \\
(2MASS) & (Gaia EDR3) & \multicolumn{1}{c}{(km s$^{-1}$)} & (kpc) & (kpc) & \multicolumn{1}{c}{(mas yr$^{-1}$)} & \multicolumn{1}{c}{(mas yr$^{-1}$)} & \multicolumn{1}{c}{(kpc)} & \multicolumn{1}{c}{(km s$^{-1}$)} \\ \midrule \midrule
\endhead
\hline
\multicolumn{9}{c}{\textit{(continued)}}\\
\endfoot
\endlastfoot

00003305-7953389 & 4634573766005607552      & 269.3     & 2.51      & 0.06      & -4.538    & -4.538    & -1.48     & 274.3      \\
00020162-4430117 & 4994519032163925632      & 58.5      & 3.13      & 0.19      & 14.656    & 14.656    & -2.92     & 202.6      \\
00040305-6106367 & 4905632480654004608      & -6.0      & 0.12      & 0.00      & 53.448    & 53.448    & -0.07     & 15.7       \\
00043646+4124062 & 384060304935385984       & -13.7     & 1.01      & 0.01      & -7.058    & -7.058    & -0.33     & 28.8       \\
00045403+3524010 & 2876804519751163008      & -91.5     & 1.06      & 0.02      & 9.525     & 9.525     & -0.45     & 154.2      \\
00052731+0025510 & 2546113345293028608      & -2.9      & 0.19      & 0.01      & -9.668    & -9.668    & -0.14     & 41.4       \\
00060902-6238019 & 4904480471642380928      & 65.5      & 5.74      & 0.41      & -4.387    & -4.387    & -4.60     & 198.3      \\
00071189+4724466 & 393031258963415936       & 13.4      & 0.15      & 0.01      & 74.884    & 74.884    & -0.01     & 74.5       \\
\bottomrule
\end{longtable}

\end{document}